\documentclass[article]{aa}
\usepackage{txfonts}
\usepackage{amssymb}
\usepackage{amsmath}
\usepackage{subcaption}
\usepackage{gensymb}
\usepackage{wasysym}
\usepackage{graphicx}
\usepackage{braket}

\usepackage{txfonts}

\begin{document}

\title{Tracing the large-scale magnetic field morphology in protoplanetary disks using molecular line polarization} 
\titlerunning{Spectral line polarization in disks}
\author{Boy Lankhaar
        \and
        Wouter Vlemmings
        \and
        Per Bjerkeli
        }
\institute{Department of Space, Earth and Environment, Chalmers University of Technology, Onsala Space Observatory, 439 92 Onsala, Sweden \\
\email{boy.lankhaar@chalmers.se}}

\date{Received ... ; accepted ...}

\abstract
   {Magnetic fields are fundamental to the accretion dynamics of protoplanetary disks and they likely affect planet formation. Typical methods to study the magnetic field morphology observe the polarization of dust or spectral lines. However, it has recently become clear that dust-polarization in ALMA's (Atacama Large (sub)Millimeter Array) spectral regime does not always faithfully trace the magnetic field structure of protoplanetary disks, which leaves spectral line polarization as a promising method for mapping the magnetic field morphologies of such sources.}
   {We aim to model the emergent polarization of different molecular lines in the ALMA wavelength regime that are excited in protoplanetary disks. We explore a variety of disk models and molecules to identify those properties that are conducive to the emergence of polarization in spectral lines and may therefore be viably used for magnetic field measurements in protoplanetary disks.}
   {We used POlarized Radiative Transfer Adapted to Lines (PORTAL) in conjunction with the Line Emission Modeling Engine (LIME). Together, they allowed us to treat the polarized line radiative transfer of complex three-dimensional physical and magnetic field structures.} 
   {We present simulations of the emergence of spectral line polarization of different molecules and molecular transitions in the ALMA wavelength regime. We find that molecules that thermalize at high densities, such as HCN, are also the most susceptible to polarization. We find that such molecules are expected to be significantly polarized in protoplanetary disks, while molecules that thermalize at low densities, such as CO, are only significantly polarized in the outer disk regions. We present the simulated polarization maps at a range of inclinations and magnetic field morphologies, and we comment on the observational feasibility of ALMA linear polarization observations of protoplanetary disks.}
{We conclude that those molecules with strong dipole moments and relatively low collision rates are most useful for magnetic field observations through line polarization measurements in high density regions such as protoplanetary disks.}

\maketitle
\section{Introduction}
The formation of solar-type stars is accompanied by a protoplanetary disk. The protoplanetary disk is a thin rotating disk structure composed of dense gas and dust that surrounds the forming (proto)star. Protoplanetary disks are believed to be sites of ongoing planet formation, and they are vital to the accretion of mass onto the central forming (proto)star \citep{armitage:19}. The dynamics of a protoplanetary disk are strongly impacted by a magnetic field \citep{li:14a}. Magnetic fields are important to many aspects of disk accretion, since they can provide the transport of angular momentum through magnetic tension, or through the launching of an outflow \citep{blandford:82, bjerkeli:16}. Magnetic fields are a key part in the magneto rotational instability (MRI) \citep{balbus:91}, which is believed to be one of the major sources of turbulence in the disk \citep{flock:17} and also key to amplifying the viscosity to allow for efficient accretion \citep{pringle:81}. To properly determine the importance of the magnetic field in disk-dynamical processes, we require direct measurements of the strength of a magnetic field, as well as the magnetic field morphology. 

Magnetic field determination may be done via polarization observations. The magnetic field strength may be derived from the circular polarization of atomic or molecular lines. Atomic and molecular lines produce circular polarization through the Zeeman effect, which is strongest for paramagnetic species with at least one unpaired electron. Magnetic fields have been shown to be present and important to the dynamics on the surface of accreting protostars through atomic line observations \citep{donati:11, sokal:18}, as well and on larger scales of star formation through molecular Zeeman observations \citep{crutcher:10, hull:19}. Zeeman observations have also been performed in the absolute inner regions ($0.05$ au) of FU Orionis, where \citet{donati:05} detected a strong magnetic field of $1$ kG. At higher distances from the protostar, \citet{vlemmings:19} attempted to perform a direct magnetic field detection in the disk of TW Hya through CN Zeeman observations using the Atacama Large (sub)Millimeter Array (ALMA). However, the (vertical component of the) magnetic field did not give rise to detectable circular polarization signals. Similar observations have been performed for AS 209, but they failed to directly detect the magnetic field strength \citep{harrison:21}. Linear polarization observations may provide information on the magnetic field morphology. In the inner $1$ au regions of the protoplanetary disk, this may be done through atomic line polarization observations. Atomic lines may linearly polarize through the ground-state-alignment (GSA) mechanism \citep{yan:06, zhang:20}, and this has also been theoretically predicted to occur in the inner regions of accretion disks \citep{yan:08}. Farther out than 1 au, in the molecular disk, where most of the disk mass resides and where planet formation is believed to be on-going, there are no direct measurements of the magnetic field strength or morphology. 

The magnetic field morphology in molecular gas is typically determined through the observation of polarized radiation from dust or molecules \citep{crutcher:19}. The polarization direction is indicative of the alignment of the emitting species. Typically, dust aligns itself with respect to the magnetic field direction via the radiative torque (RAT) mechanism \citep{lazarian:07}, while molecules align themselves with respect to the magnetic field through the Goldreich-Kylafis (GK) effect \citep{goldreich:81}. It has recently become clear that the RAT alignment mechanism competes with alternative alignment mechanisms such as self-scattering or radiative alignment in protoplanetary disks and that, therefore, the polarized emission of dust does not faithfully trace the magnetic field direction \citep{kataoka:15, kataoka:17, stephens:17}. On the other hand, the line-emission polarization angle due to the GK effect has a $90^o$ ambiguity with respect to the projected magnetic field direction, and polarization of spectral lines will only manifest in regions where radiative interactions are strong and anisotropic \citep{goldreich:81, lankhaar:20a}.

The GK effect is a consequence of the different absorption (or stimulated emission) probabilities of $\Delta m = \pm 1$ ($\sigma^{\pm}$) and $\Delta m = 0$ ($\pi$) transitions when resonant radiation is anisotropic \citep{goldreich:81}. When either $\sigma^{\pm}$ or $\pi$ transitions are favored, this causes an overpopulation of some magnetic sublevels with respect to others. Generally, collisions do not favor the population of any magnetic sublevel, and they therefore tend to isotropize the magnetic sublevels (see \citet{goldreich:81} and \citet{landi:06}. For an exception to this, see \citet{lankhaar:20b}). When the magnetic substates within a rotational level are differently populated, then this rotational level can be thought of as being aligned \citep{blum:81,landi:06, lankhaar:20a}. If, furthermore, the magnetic precession rate (on the order of s$^{-1}$/mG for a nonparamagnetic molecule) is higher than collisional or radiative interaction rates (typically $\gtrsim 10^{-4}$ s$^{-1}$), then the alignment is either parallel or perpendicular to the magnetic field direction \citep{landi:06}. The alignment of an energy level causes transitions that involve that energy level to partially polarize in the direction of the alignment of the participating energy levels. This process, the GK effect, has been detected in star-forming regions \citep{cortes:05, cortes:06}, outflows \citep{glenn:97, girart:99, lee:18}, and in the circumstellar envelope around evolved stars \citep{glenn:97b, girart:12, vlemmings:12, vlemmings:17, huang:20}, and it has provided observers with information on the magnetic field morphology of the investigated sources. But the GK effect has yet to be detected in protoplanetary disks \citep{stephens:20}. 

Numerical modeling of GK effects has traditionally been done using the large velocity gradient (LVG) approximation \citep{goldreich:81, deguchi:84, cortes:05}. LVG modeling finds that polarization is maximal for optical depths around unity and for strong (compared to collisional) radiative interactions. However, the LVG approximation is not particularly suited to model three-dimensional and direction-dependent properties of a protoplanetary disk due to its complex geometry. Instead, in order to properly interpret and predict polarization signals due to the GK effect in a complicated physical structure such as a protoplanetary disk, we have to perform three-dimensional (3D) polarized line-radiative transfer simulations. Recently, we developed a 3D POlarized Radiative Transfer code Adapted to Lines (PORTAL), which is presented in \citet{lankhaar:20a}. With PORTAL, the emergence of polarized line emission in geometrically complex astrophysical structures and magnetic fields can be investigated. PORTAL can treat the extensive energy-level structure of a number of molecules, also including interactions with vibrationally excited states. 

In this paper, we explore the polarization properties of different molecules in protoplanetary disks with a range of magnetic field configurations. We investigate the influence of disk characteristics such as the disk mass and velocity profile on the polarization characteristics of molecular lines. We  particularly focus on CO, the most abundant molecule after H$_2$ in protoplanetary disks, and HCN, which has a particularly strong dipole moment, thought to be conducive to the emergence of polarization in its spectral lines. In section 2, we briefly summarize the characteristics of PORTAL and the LIne Modeling Engine (LIME), which we have used in our simulations, and we outline the relevant characteristics of the disk model we have used, as well as the molecular excitation modeling. In section 3, we present and comment on the simulations of a fiducial disk model. We present the polarization characteristics of the spectral lines of CO and HCN for different inclinations. In section 4, we comment on the effect of variations on the fiducial disk model on the polarization characteristics of spectral lines excited in protoplanetary disks. We end the discussion section by commenting on the observational feasibility of linear polarization observations in protoplanetary disks. In section 5, we summarize our findings.
\section{Methods}
We simulated the emergence of polarization in molecular spectral lines in protoplanetary disks using PORTAL. PORTAL is a parallelized 3D polarized radiative transfer code that is adapted to spectral lines. PORTAL does not assume local thermal equillibrium (LTE) excitation. PORTAL uses the anisotropic intensity and strong magnetic field approximations, which are explained in \citet{lankhaar:20a}. As its input, PORTAL requires a physical structure, including gas density, temperature, and other parameters relevant to the molecular excitation. With that information, PORTAL performs a comprehensive ray tracing of the radiation environment throughout the simulation. Specifically, PORTAL computes, for all molecular transitions, $i \to j$, with resonant frequencies $\nu_{ij}$, the total integrated intensity, $J_0^0(\nu_{ij}) = \int d\Omega \int d\nu \ I(\nu,\Omega) \phi(\nu-\nu_{ij})$, as well as the anisotropic intensity $J_0^2(\nu_{ij}) =\int d\Omega \ P_2 (\cos \theta )/\sqrt{2} \int d\nu \ I (\nu,\Omega) \phi (\nu-\nu_{ij})$. In the previous equations, $I(\nu,\Omega)$ is the radiation specific intensity in direction $\Omega$ and frequency $\nu$, and $\phi (\nu-\nu_{ij})$ is a line profile around $\nu_{ij}$. The anisotropic intensity is with respect to the magnetic field direction ($\cos \theta = \hat{\boldsymbol{k}}_{\Omega} \cdot \hat{\boldsymbol{b}}$, where $\hat{\boldsymbol{k}}_{\Omega}$ and $\hat{\boldsymbol{b}}$ are unit vectors with respect to the propagation and magnetic field direction), and weighted by the second-order Legendre polynomial $P_2 (\cos \theta )$. The radiation field solution is subsequently used to compute the quantum state alignment throughout the simulation by solving the polarized statistical equilibrium equations given in equation 5 of \citet{lankhaar:20a}. The polarized excitation solution obtained by PORTAL may then be used to ray trace a polarized image of the investigated astrophysical regions by solving the polarized radiative transfer equation given by equations 6-9 of \citet{lankhaar:20a}. Radiation sources such as a star may be added to the simulation. PORTAL can be used in a stand-alone mode, assuming LTE excitation, but in this work, we used the molecular excitation solutions of LIME\footnote{We used LIME version 1.9.5, available from \url{https://github.com/lime-rt/lime}. We made minor adjustments to the output of LIME to make it compatible with PORTAL.} in conjunction with PORTAL\footnote{The source code of PORTAL is available on GitHub at \url{https://github.com/blankhaar/PORTAL}.}. 

LIME is a parallelized non-LTE 3D line radiative transfer code that uses a random grid connected through a Delaunay tessellation \citep{brinch:10}. Arbitrary physical structures and geometries may be input. With the input of a physical structure, LIME computes the radiative transfer and molecular excitation using a Monte-Carlo scheme, which is sped up by an accelerated lambda iteration \citep{rybicki:91}. 

In the following, we briefly introduce the protoplanetary disk physical model that we use. We performed simulations on a range of molecules and before presenting the results of the simulations, we discuss some general characteristics of these molecules that are relevant to their polarization.

\begin{table*}[t]
\caption{Properties of the molecular transitions that we investigate in this work. We report the frequency in GHz, molecular abundance, and critical density in cm$^{-3}$ of the molecular transitions.}
\centering
\begin{tabular}{l l c c c}
\hline \hline 
Molecule & Transition & $\nu_0$ (GHz) & $x_{\mathrm{mol}}$ & $n_{\mathrm{crit}}$ (cm$^{-3}$) \\
\hline
CO     & $J=3 \to 2$ & $345.796$ & $10^{-4}$ &  $1.75\times10^{4}$ \\
CS     & $J=8\to 7 $ & $342.883$ & $10^{-9}$ &  $7.75\times10^{6}$ \\
HCN    & $J=4\to 3 $ & $354.505$ & $10^{-8}$ &  $1.88\times10^{7}$ \\
HCO$^+$& $J=4\to 3 $ & $356.734$ & $10^{-9}$ &  $4.36\times10^{6}$ \\
N2H$^+$& $J=4\to 3 $ & $372.673$ & $10^{-10}$ & $3.71\times10^{6}$ \\ 
\hline
\end{tabular}
\label{tab:AoC}
\end{table*}

\subsection{Protoplanetary disk model}
We assumed an axisymmetric tapered disk as our protoplanetary disk model \citep{andrews:09, lynden-bell:74, hartmann:98}. The tapered disk model of the density structure, $\rho(r_c,z)$, is dependent on the parameters $\gamma$, the characteristic radius $R_c$, and a scale-height profile defined by the scale height at $100$ au, $H_{100 \ \mathrm{au}}$, and $\psi$: 
\begin{subequations}
\label{eq:density}
\begin{align}
\rho (r_c, z) &= \frac{\Sigma (r_c)}{\sqrt{2\pi} H} e^{-(z/H)^2/2}, \\
\Sigma (r_c) &= \Sigma_c \left(\frac{r_c}{R_c}\right)^{-\gamma} e^{-(r_c/R_c)^{2-\gamma}}, \\
H &= H_{100 \ \mathrm{au}} \left(\frac{r_c}{100 \ \mathrm{au}}\right)^{1 + \psi},
\end{align}
where $r_c=\sqrt{x^2+y^2}$ is the cylindrical radius and $z$ is the height. The total disk mass can be related to the characteristic column density when $\gamma \neq 2$, via $\Sigma_c = \frac{(2-\gamma)M_{\mathrm{disk}}}{2 \pi R_c^2}$. To relate the density to the number density, we assumed a mean molecular mass of $2.4$ hydrogen masses. We represented the gas temperature of the disk by a power law, with an additional vertical profile characterized by the atmosphere temperature, $T_{\mathrm{atm}}$, and the midplane temperature, $T_{\mathrm{mid}}$ \citep{huang:18}. For $z < 4H$, we used the temperature
\begin{align}
T(r_c,z) &= T_{\mathrm{atm}} \left(\frac{r_c}{10 \ \mathrm{au}}\right)^{-q} (1 - \cos^4[\pi z / 8 H]) \\ \nonumber 
&+ T_{\mathrm{mid}} \left(\frac{r_c}{10 \ \mathrm{au}}\right)^{-q} \cos^4[\pi z / 8 H],
\end{align}
\end{subequations}
while for $z \geq 4H$, we used $T(r_c) =  T_{\mathrm{atm}} \left(\frac{r_c}{10 \ \mathrm{au}}\right)^{-q}$. The velocity structure of the gas was assumed to be Keplerian motion around the central protostar. 

We explored the polarization properties of spectral lines under the influence of a radial, toroidal, and poloidal magnetic field configuration. We chose for the toroidal
magnetic field direction, $\boldsymbol{b}_{\mathrm{tor}}$, to be parallel to the Keplerian rotation velocity, while the poloidal magnetic field direction was chosen to be perpendicular to the toroidal and radial direction, along $\boldsymbol{b}_{\mathrm{tor}} \times \boldsymbol{b}_{\mathrm{rad}}$. 

For our fiducial model, we performed simulations on a disk with $M_{\mathrm{disk}} = 10 \ M_{\mathrm{J}}$, $\gamma = 0.5$, $R_c = 50 \ \mathrm{au}$, $H_{100 \ \mathrm{au}} = 20 \ \mathrm{au}$, $\psi = 0.28$, $q=0.5$, $T_{\mathrm{atm}} = 126.5 \ \mathrm{K}$, and $T_{\mathrm{mid}} = 31.6 \ \mathrm{K}$. For the central protostellar mass, temperature, and radius, we chose $0.5 \ M_{\odot}$, $2500$ K, and $2 R_{\odot}$, respectively. In the discussion section, we examine the effects of changing these disk parameters on spectral polarization signals. 
\subsection{Molecules}
We investigated the emergence of polarization in the spectral lines of a range of molecules. We investigated the ALMA band 7 transitions of CO, CS, HCN, HCO$^+$, and N$_2$H$^+$, of which some relevant properties are summarized in table \ref{tab:AoC}. We dedicate most of our attention to CO and HCN: CO is the second most abundant molecule, after H$_2$, and it emits strongly, while HCN is a prototypical molecule that is excited under non-LTE conditions. 



Since our analysis crucially depends on the lines being excited under non-LTE conditions, we comprehensively modeled radiative and collisional interactions in the excitation analysis. We used collisional and radiative rate coefficients from the LAMDA database \citep{schoier:05}. The collisional rate coefficients for CO come from \citet{yang:10}, for HCN they come from \citet{dumouchel:10}, for HCO$^+$ they come from \citet{flower:99}, and for CS they come from \citet{lique:06}. We use the collisional rate coefficients of HCO$^+$ for N$_2$H$^+$, as is standard in the LAMDA database, and this is warranted because of their similar collision dynamics. When densities exceed the critical density by a factor of $10-100$, LTE excitation is attained and alignment is effectively quenched. In table \ref{tab:AoC}, we list the critical densities of the lines under investigation. We computed the critical densities using equation 11 of \citet{pavlyuchenkov:07}. 

The formalism of \citet{lankhaar:20a} relies on the assumption that the magnetic precession rate is comfortably ($10-100$ times) higher than the radiative or collisional interaction rates. To confirm this is the case for the molecules under investigation, we compared the magnetic precession rates, $g\Omega = \bar{g}\mu_N / \hbar = 4.8 \bar{g} \ \mathrm{s}^{-1}/\mathrm{mG}$, where $\bar{g}$ is the transition-specific g-factor, to the transition Einstein A-coefficient. For CO and CS, we only have a contribution from the molecular rotation to the magnetic precession. The transition g-factors \citep{flygare:71} are $g_{\mathrm{CO}}=0.269$ and $g_{\mathrm{CS}}=0.272$, leading to ratios, $\Gamma_{\mathrm{mag}}^{1 \ \mathrm{mG}} = \frac{g\Omega_{1 \ \mathrm{mG}}}{A_{ul}}$, of $\left(\Gamma_{\mathrm{mag}}^{1 \ \mathrm{mG}}\right)_{\mathrm{CO}}=5.0\times 10^5$ and $\left(\Gamma_{\mathrm{mag}}^{1 \ \mathrm{mG}}\right)_{\mathrm{CS}}=1.6\times 10^3$. So magnetic fields of $\gtrsim 0.02 \ \mathrm{\mu G}$ and $\gtrsim 6.3 \ \mathrm{\mu G}$ ensure magnetic alignment over radiative alignment for CO and CS. The molecules HCN, HCO$^+$, and N$_2$H$^+$ have contributions from both their nuclear spin magnetic moment as well as their rotational magnetic moment to the magnetic precession rate. The level specific g-factor that defines the proportionality of the magnetic precession rate with the magnetic field is called the Land\'{e} g-factor. The contribution of the nuclear spin magnetic moment to the total magnetic precession may be high for low-J rotational states. This is in particular the case for non-LTE excited molecules, which are not expected to host evenly distributed hyperfine states \citep{lankhaar:18, keto:10}. We assume the Land\'{e} g-factor of HCN, HCO$^+$, and N$_2$H$^+$ around $g_l=0.1$, which puts their characteristic ratios at
$\left(\Gamma_{\mathrm{mag}}^{1 \ \mathrm{mG}}\right)_{\mathrm{HCN}}=2.3 \times 10^2$, 
$\left(\Gamma_{\mathrm{mag}}^{1 \ \mathrm{mG}}\right)_{\mathrm{HCO}^+}=1.3 \times 10^2$, and 
$\left(\Gamma_{\mathrm{mag}}^{1 \ \mathrm{mG}}\right)_{\mathrm{N}_2\mathrm{H}^+}=1.6\times 10^2$. That means that magnetic fields of $\gtrsim 43 \ \mathrm{\mu G}$, $\gtrsim 77 \ \mathrm{\mu G}$, and $\gtrsim 63 \ \mathrm{\mu G}$ ensure magnetic alignment over radiative alignment for HCN, HCO$^+,$ and N$_2$H$^+$. We also require the collisional rate, $k_{\mathrm{col}}$, to be lower than the magnetic precession rate. An order of magnitude calculation puts that ratio at $\Gamma_{\mathrm{col}} = \frac{g\Omega}{k_{\mathrm{col}}} \sim 42 \bar{g} \left( \frac{B}{1 \ \mathrm{mG}} \right) \left(\frac{\sigma}{10 \ \AA^2}\right)^{-1} \left( \frac{n}{10^9 \mathrm{cm}^{-3}} \right)^{-1} \left( \frac{T}{100 \ \mathrm{K}} \right)^{-\frac{1}{2}}$, where $\sigma$ is the collision cross section, $T$ the temperature, and $n$ the number density. The above estimates warrant the strong magnetic field approximation for the disk molecules under investigation. 


We investigated the band 7 transitions of the molecules listed in table \ref{tab:AoC}. We focused on these transitions because (i) rates of radiative interaction increase with the transition frequency with respect to collisional rates and (ii) the spectral flux densities are comparatively stronger for these lines in the regions we are interested in. For these reasons, we expect high polarization fractions, as well as high polarized flux densities for the band 7 transitions. We make a comparison to lower frequency transitions in section 4.3.1. We focused on CO and HCN as the main molecules of investigation. It should be noted from table \ref{tab:AoC} that CO transitions have comparatively low critical densities, meaning that their radiative interactions are relatively weak compared to collisional interactions. Indeed, when analyzing spectral radiation emitted from protoplanetary disk CO, it is often assumed that the energy levels are Boltzmann distributed. It should, however, also be stressed that we expect CO emission to trace regions that are relatively high up in the atmosphere of the disk, where the gas is more diffuse \citep{zhang:17}. The strong dipole moment of HCN leads to relatively strong radiative interactions, which means that its transitions are susceptible to polarize, provided the radiation morphology is anisotropic. Since the isotopologues of CO are observed in protoplanetary disks as well, we also investigated their polarization properties (see section 4.3.2).

Chemical modeling of protoplanetary disks shows that the molecular abundance profile throughout the disk varies tremendously, through freeze-out processes, UV dissociation, and other chemical reactions \citep{walsh:10, walsh:12}. We did not include spatial variation of the molecular abundance in our modeling, but we do discuss possible effects of it on the polarization signal in section 4.2. 

\begin{figure}[h!]
  \centering
  \begin{subfigure}[b]{0.45\textwidth}
    \includegraphics[width=\textwidth]{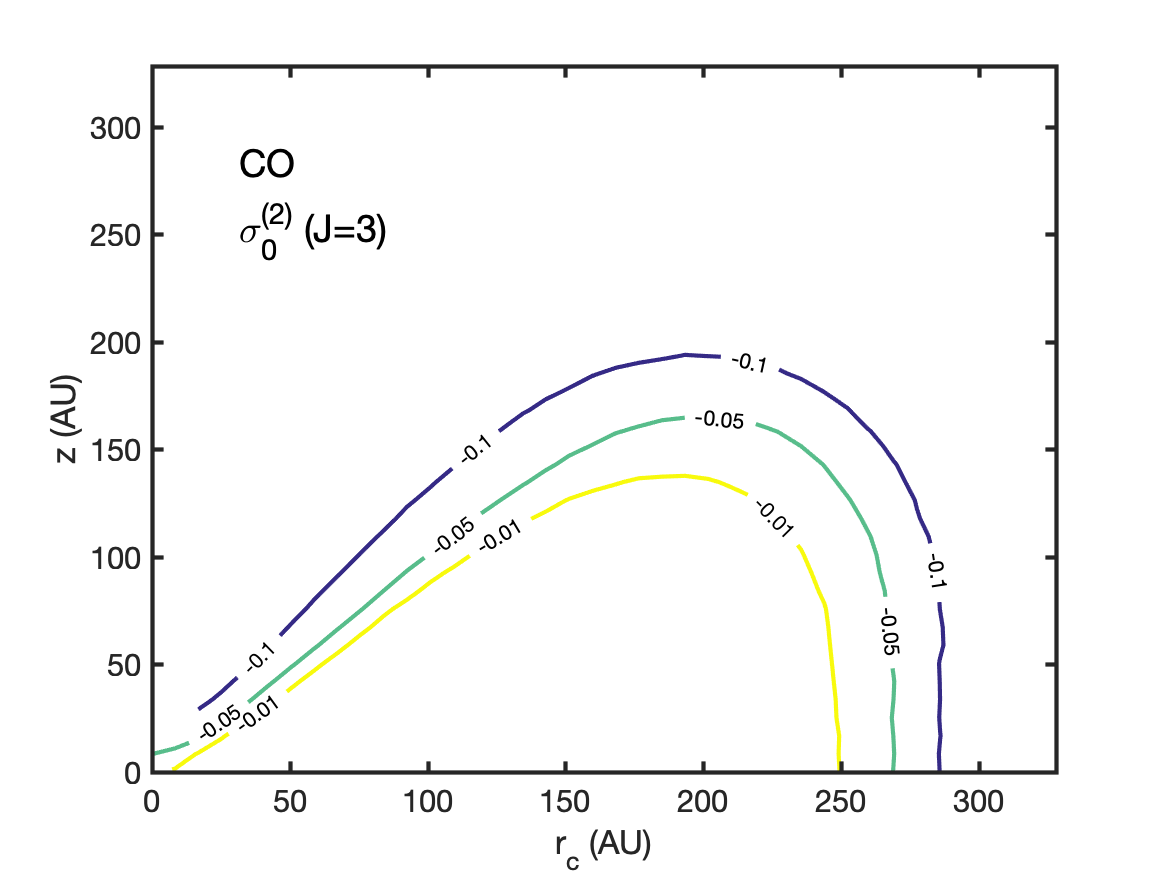}
    \caption{}
    \label{fig:pp_s2_CO}
  \end{subfigure}
  ~
  \begin{subfigure}[b]{0.45\textwidth}
    \includegraphics[width=\textwidth]{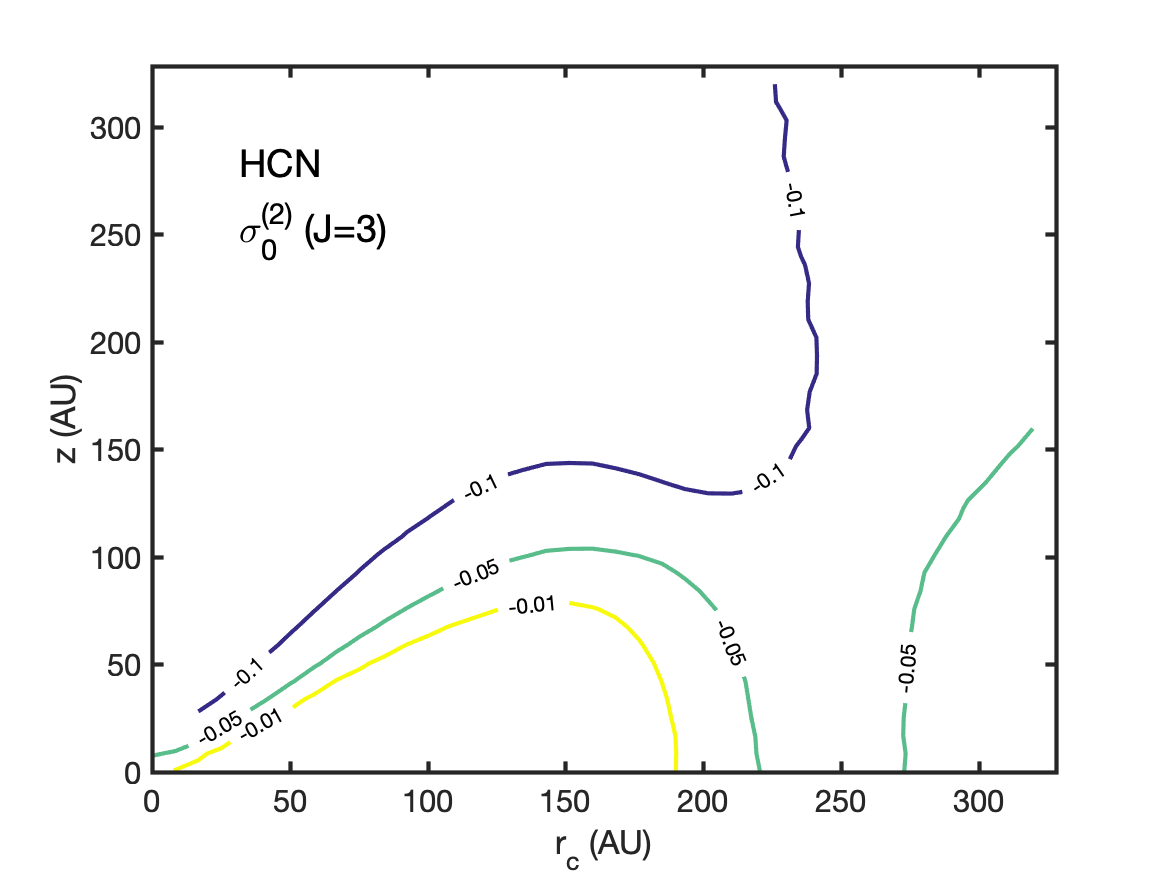}
    \caption{}
    \label{fig:pp_s2_HCN}
  \end{subfigure}
  \caption{Relative alignment of the $J=3$-state of the molecules CO (a) and HCN (b). The relative alignment was computed assuming a toroidal magnetic field.}
  \label{fig:rel_all}
\end{figure}
\section{Simulations}
We present our PORTAL simulation results by first discussing the results for our fiducial protoplanetary disk system. We present the characteristics of the alignment for CO and HCN, and show polarization maps of their $J=3\to2 $ and $J=4\to3$ transitions at different inclinations. After that, in the discussion section, we examine the influence of some of the disk parameters on the alignment and polarization properties. 

We begin our discussion of the PORTAL simulations by analyzing the relative alignment of the molecular states, which are represented by the parameter $\sigma_0^{(2)} (j)$ \citep{lankhaar:20a}. In Fig.~\ref{fig:rel_all}, we plot the relative alignment of the $J=3$ states of CO and HCN, for an axisymmetric disk as a function of the cylindrical radius and the height. We find that significant alignment manifests itself in the molecular states in the outer atmosphere of the disk, but that near the midplane, alignment is significantly quenched. The same trend can be seen for the relative anisotropy of the radiation. Indeed, we expect that both the density and optical depth increase toward the midplane. Collisions are more prominent in the denser regions of the disk, while a high optical depth tends to isotropize the radiation field. Both of these factors result in no significant alignment of the quantum states, and accordingly, no polarization in the associated transitions. In the disk atmosphere and outer regions of the disk, optical depths and densities become lower, so that alignment and anisotropy can manifest itself in the molecular states and radiation field.

Comparing the alignment properties of different disk molecules, we observe that CO only gets significantly aligned farther out in the atmosphere and the disk, when compared to HCN. This is to be expected due to the low critical density of CO. We note that HCN thermalizes at densities of about 2 orders of magnitude higher than CO. Collisional interactions thus dominate and quench alignment in CO already at far lower densities. The other molecules that we investigated, HCO$^+$, CS, and N$_2$H$^+$, have alignment properties somewhere between the extremes of CO and HCN. Particularly, the molecular ions have a similar dipole moment to HCN, but relatively strong collisional interactions. We assumed the collisional interactions of molecular ions to be isotropic, but they might be anisotropic due to an ambipolar velocity drift \citep{lankhaar:20b}, which would provide additional alignment.

\begin{figure*}[h!]
  \centering
  \begin{subfigure}[b]{0.31\textwidth}
    \includegraphics[width=\textwidth]{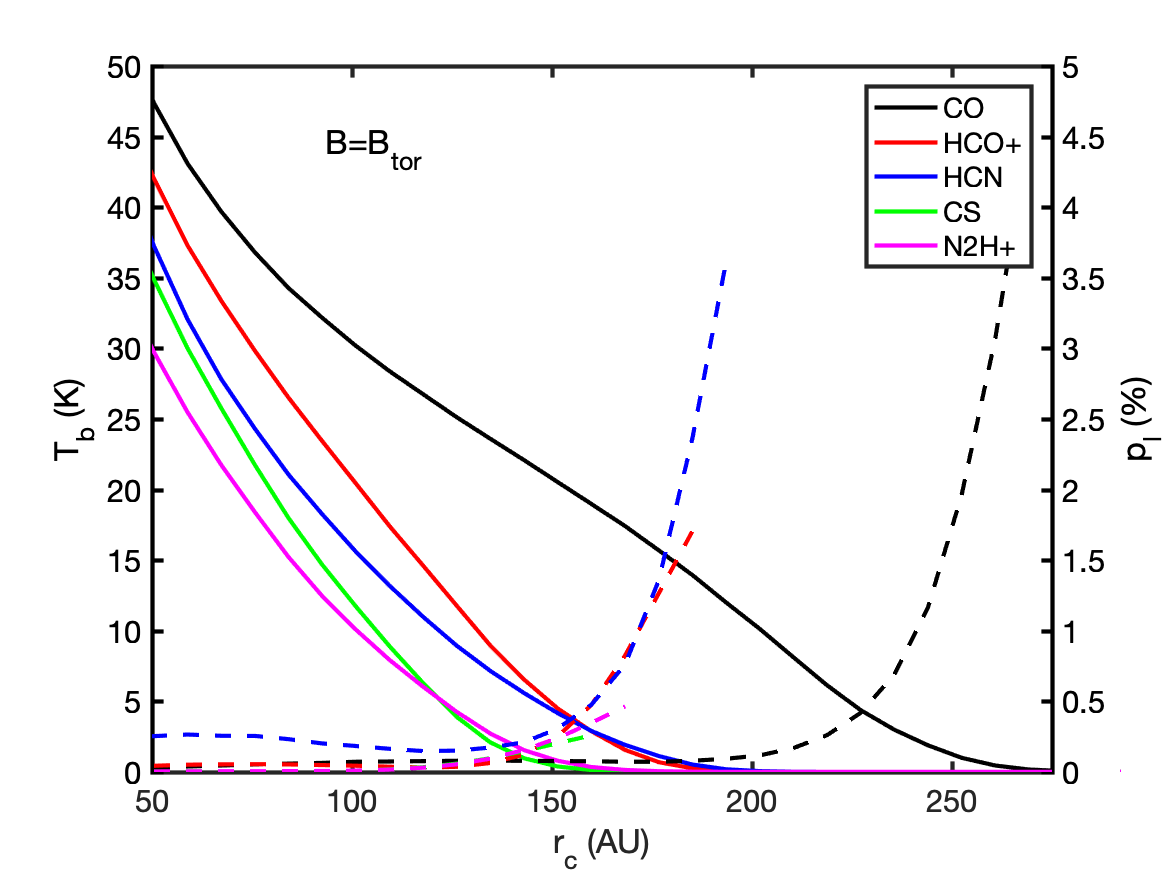}
    \caption{}
    \label{fig:spec_face_tor}
  \end{subfigure}
  ~
  \begin{subfigure}[b]{0.31\textwidth}
    \includegraphics[width=\textwidth]{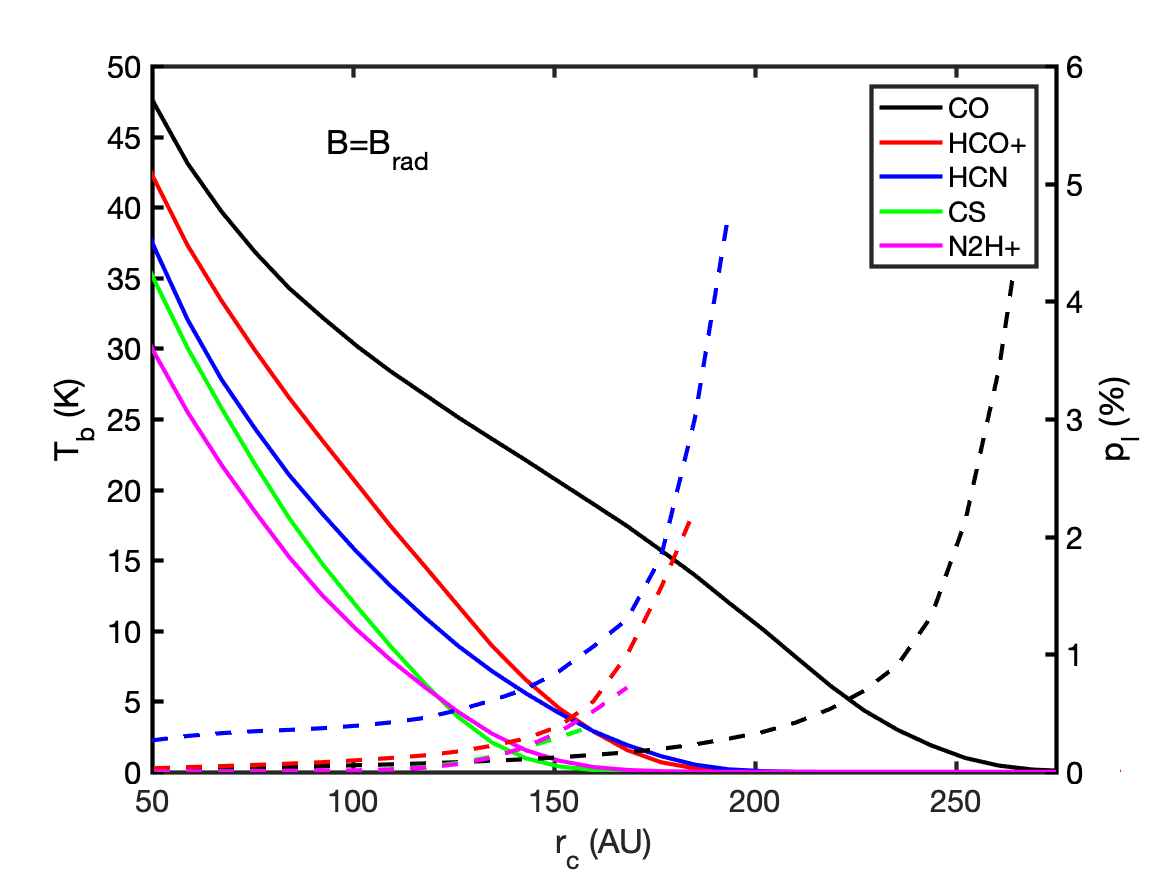}
    \caption{}
    \label{fig:spec_face_rad}
  \end{subfigure}
  ~
  \begin{subfigure}[b]{0.31\textwidth}
    \includegraphics[width=\textwidth]{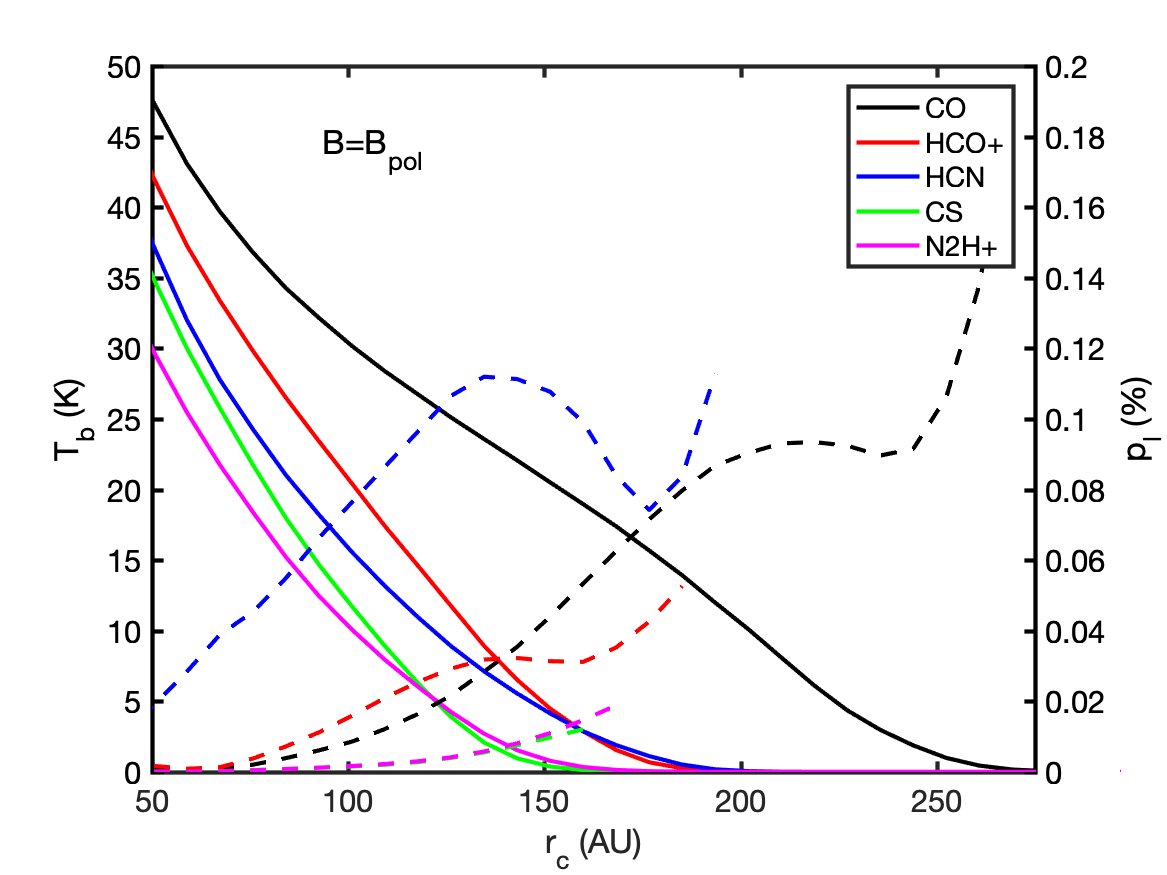}
    \caption{}
    \label{fig:spec_face_pol}
  \end{subfigure}
  \caption{Azimuthally averaged total emission and polarization fraction of transitions of a range of molecules from a face-on axisymmetric disk. The relevant parameters for the molecular transitions are given in Table \ref{tab:AoC}. The polarization fraction is omitted for $T_b < 0.1$ K.}
  \label{fig:spec_face_all}
\end{figure*}

We move on to estimates of the polarization fraction for different molecules from a face-on ($i=0^o$) image of the protoplanetary disk. We plotted polarization fraction estimates for the ALMA band 7 transitions of CO, HCO$^+$, HCN, CS, and N$_2$H$^+$, coming from a face-on ($i=0^o$) oriented protoplanetary disk in Fig.~\ref{fig:spec_face_all}. Because images of a face-on disk are axisymmetric (and we explore only axisymmetric magnetic fields), we averaged them azimuthally and plotted the brightness temperature and polarization fraction as a function of the projected distance to the central star. We considered the emergent polarization fraction for the following three different magnetic field configurations: a toroidal, poloidal, and radial magnetic field. 
\begin{figure*}[h!]
  \centering
  \begin{subfigure}[b]{0.9\textwidth}
    \includegraphics[width=\textwidth]{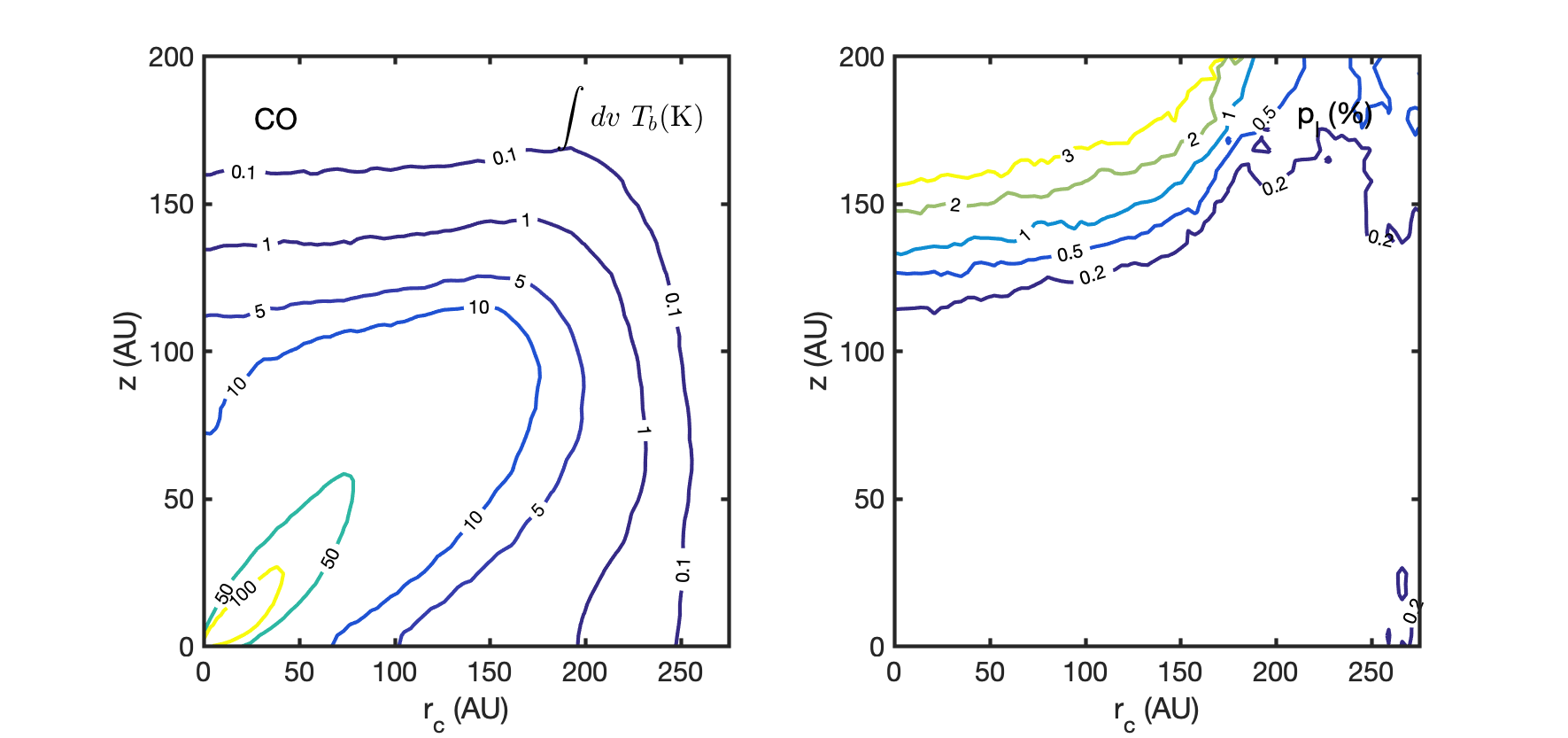}
    \caption{}
    \label{fig:edge_int_CO}
  \end{subfigure}
  ~
  \begin{subfigure}[b]{0.9\textwidth}
    \includegraphics[width=\textwidth]{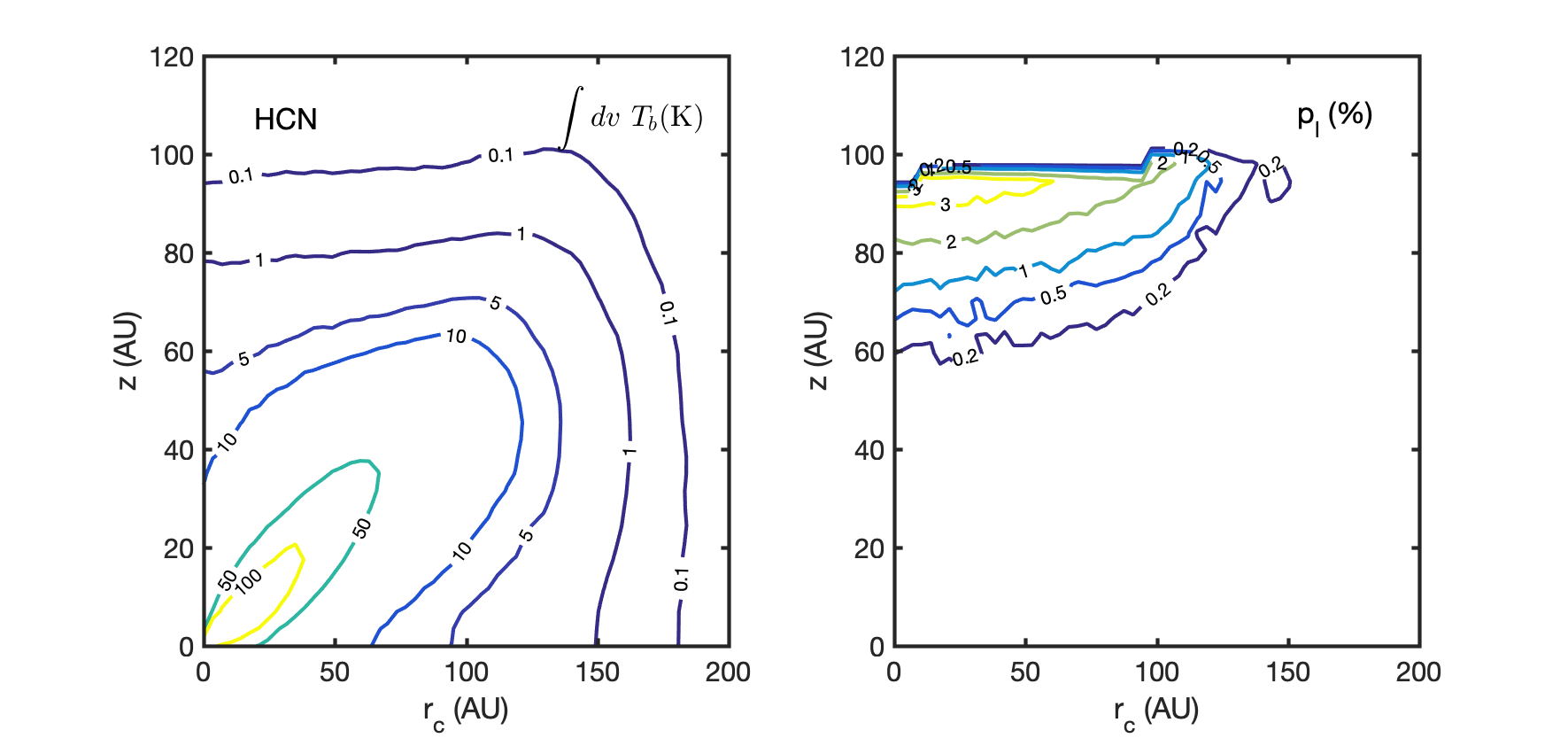}
    \caption{}
    \label{fig:edge_int_HCN}
  \end{subfigure}
  \caption{Integrated intensity maps of an edge-on protoplanetary disk (left side) and the associated polarization fraction at the maximum intensity at that position (right side). We plotted the results for the CO $J=3-2$ transition (a) and the $J=4-3$ transition of HCN (b). The magnetic field has a toroidal morphology.}
  \label{fig:edge_int}
\end{figure*}

As expected, brightness temperatures are high close to the central protostar and decrease toward the outer parts of the disk. Also, CO emission is more extended compared to the other species. For all species and magnetic field morphologies, we see a strong rise in polarization in the region where the emission dies out. This can be partly ascribed to the low densities, but also to the increasing anisotropy in the radiation field in these optically thin regions. For a poloidal magnetic field morphology, we obtain weak polarization signals, which should be ascribed to the magnetic field being parallel to the propagation direction in the regions where emission is being produced\footnote{The polarized emission and absorption coefficients, $\epsilon_Q, \eta_Q \propto \sin^2 \chi$, where $\chi$ is the angle between the propagation direction and the magnetic field direction. See also equation~(8b) of \citet{lankhaar:20a}}. For both the radial and toroidal magnetic field configuration, polarization is produced in excess of half a percent of the total emission toward the outer parts of the disk. The polarization estimates of the radial magnetic field configuration are slightly higher. For all molecules, the polarization fractions increase in the outer parts of the disk where the emission dies out. Furthermore, HCN emission is polarized at around half a percent of the total emission throughout the disk, with a sharp increase in the polarization fraction in the outer parts of the disk.

\begin{figure*}[h!]
  \begin{subfigure}[b]{0.9\textwidth}
    \includegraphics[width=\textwidth]{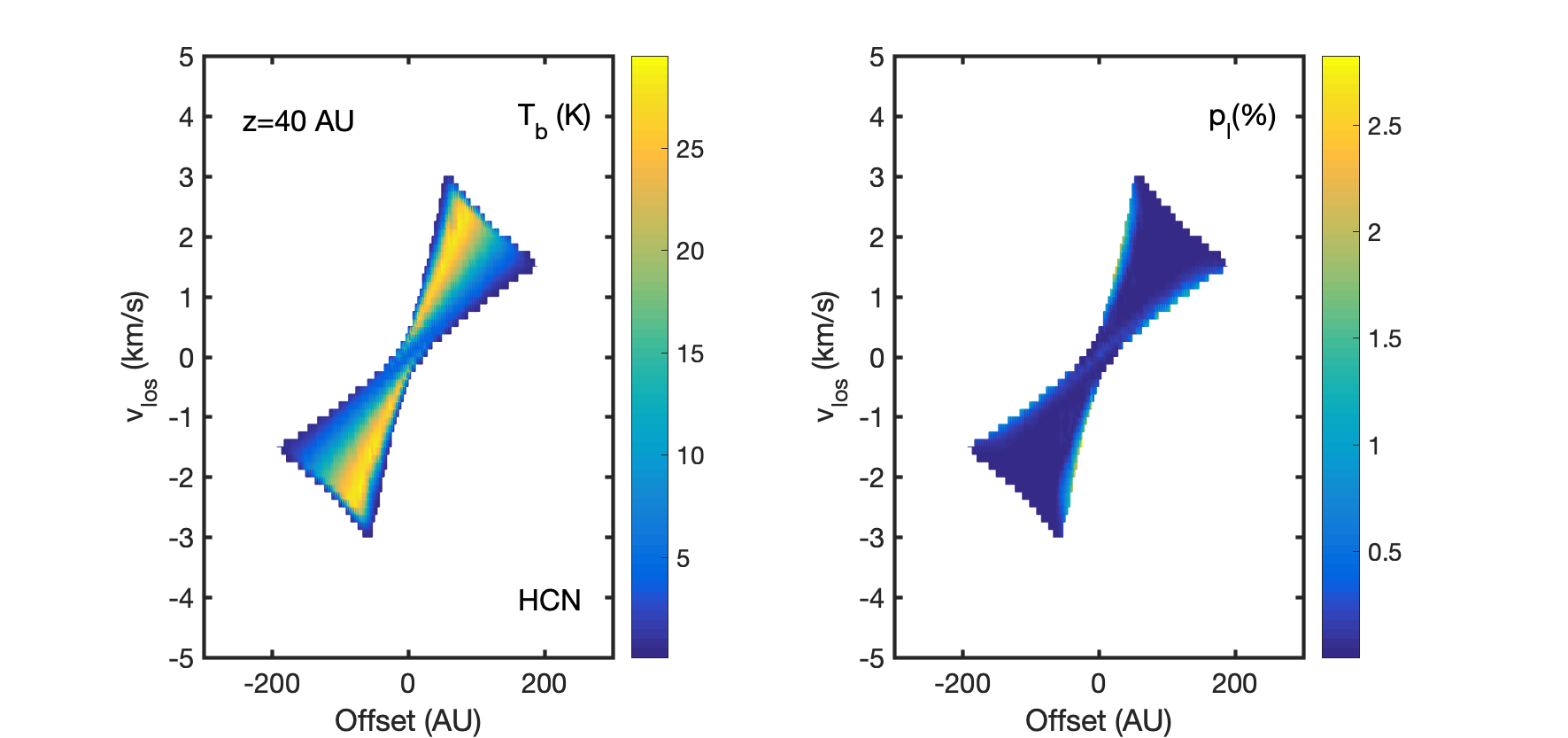}
    \caption{}
    \label{fig:edge_pv1}
  \end{subfigure}
  ~
  \begin{subfigure}[b]{0.9\textwidth}
    \includegraphics[width=\textwidth]{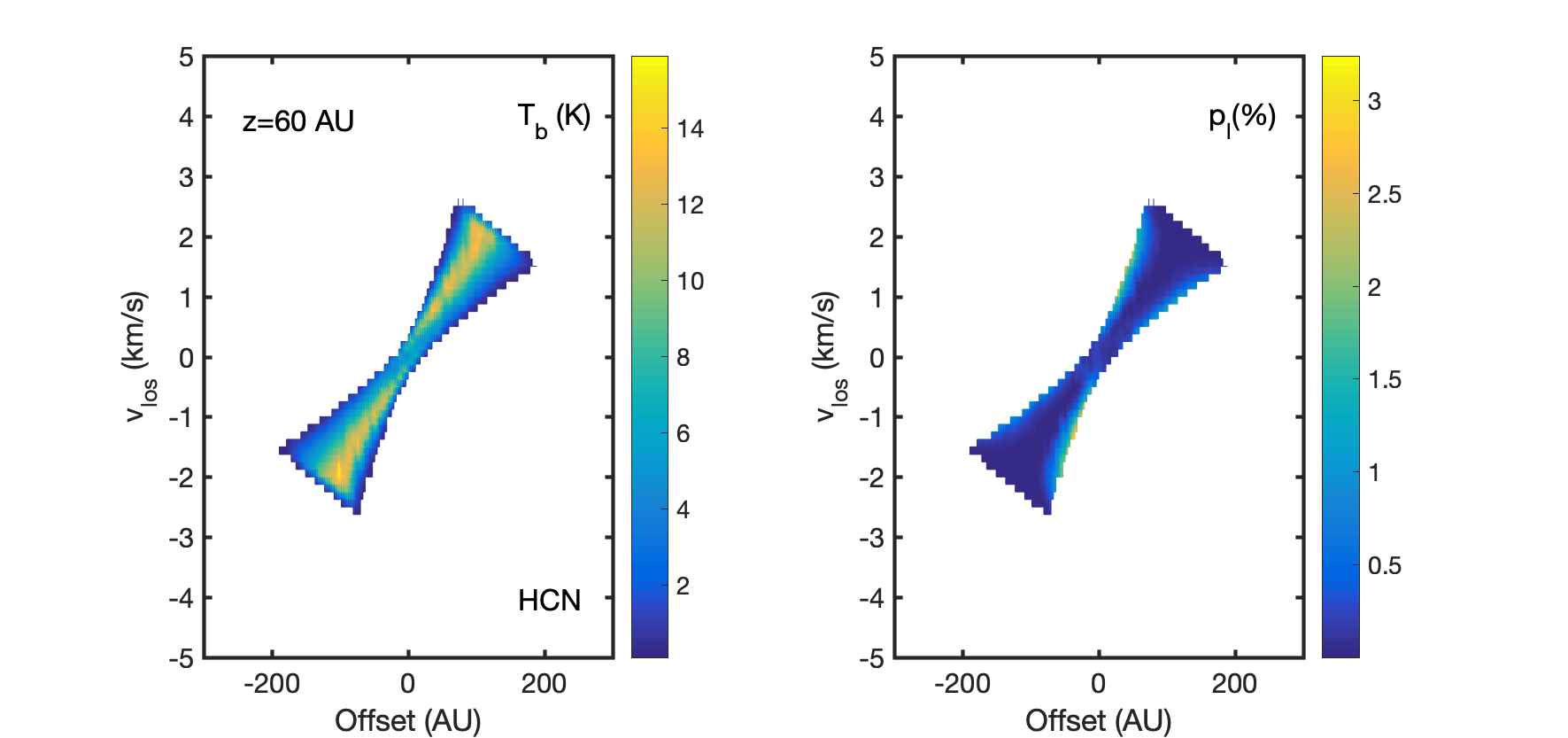}
    \caption{}
    \label{fig:edge_pv2}
  \end{subfigure}
  \caption{Position-velocity diagrams of the HCN $J=4-3$ transition for $z=40$ au (a) and $z=60$ au (b). For each height slice, we give the total brightness temperature (left) and the polarization fraction (right). The magnetic field has a toroidal morphology.}
  \label{fig:edge_pv}
\end{figure*}

We explore the polarization of molecular spectral lines in an edge-on disk in Figs.~\ref{fig:edge_int} and \ref{fig:edge_pv}. In Fig.~\ref{fig:edge_int} we report the integrated intensity maps of CO and HCN for an edge-on disk with the associated polarization fractions at the velocity channel of maximum intensity. We used a toroidal magnetic field for the edge-on disk polarization maps. Polarization fractions only exceed $0.2\%$ for a high vertical offset and low horizontal offset from the central star. In the brightest regions, emission tends to be unpolarized due to the high densities the CO emission is tracing here. Polarized emission is seen mainly at a high vertical offset at $z>100 \ \mathrm{au}$, where the diffuse disk atmosphere is traced. Accordingly, these regions are associated with a low integrated intensity $\lesssim 1  \ \mathrm{K \ km \ s}^{-1}$. We find a similar polarization morphology for HCN, but the HCN emission is polarized in excess of $0.2\%$ closer to the inner regions of the disk. Polarization fractions of $>0.2\%$ emerge in the line emission of HCN at a vertical offset $z>60 \ \mathrm{au}$, and a low horizontal offset $r_c < 150 \ \mathrm{au}$. At a large horizontal offset, polarization fractions drop because the magnetic field in these regions is along the propagation direction \citep[see also][]{lankhaar:20a}. 

In Fig.~\ref{fig:edge_pv}, we present position-velocity (PV) diagrams of the HCN emission and a polarization fraction from a toroidal magnetic field at $z=40 \ \mathrm{au}$ and $z=60\ \mathrm{au}$. In the PV diagrams, we find that polarization fractions can be very high ($>1.5\%$) in particular regions. However, particularly regions of low emission are polarized most strongly. Low polarization fractions are predicted close to the central star because these regions are the densest. At the wings of the emission, at high velocity and position offset, the highest polarization fractions are predicted.  

\begin{figure*}[h!]
  \centering
  \begin{subfigure}[b]{0.9\textwidth}
    \includegraphics[width=\textwidth]{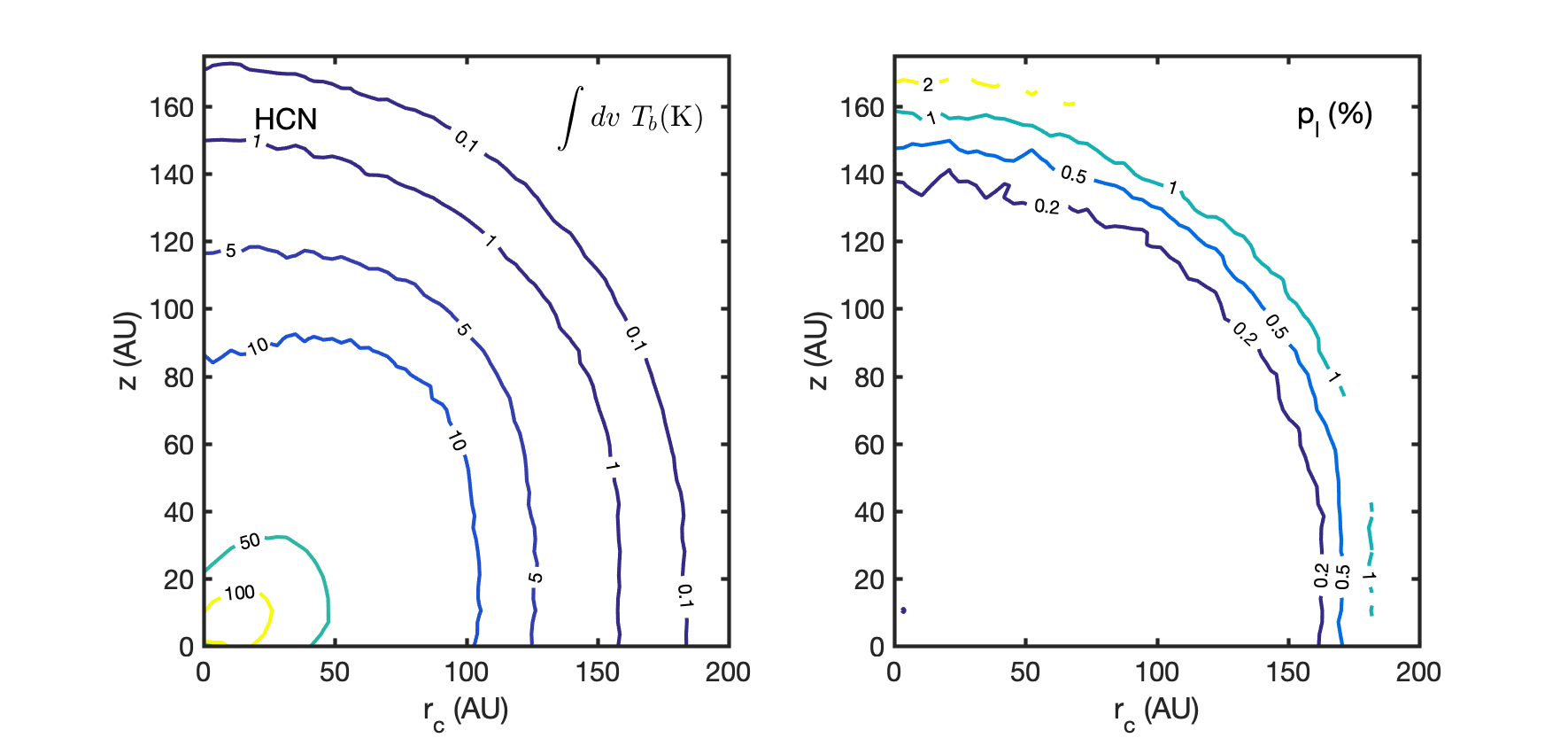}
    \caption{}
    \label{fig:45_HCN}
  \end{subfigure}
  ~
  \begin{subfigure}[b]{0.9\textwidth}
    \includegraphics[width=\textwidth]{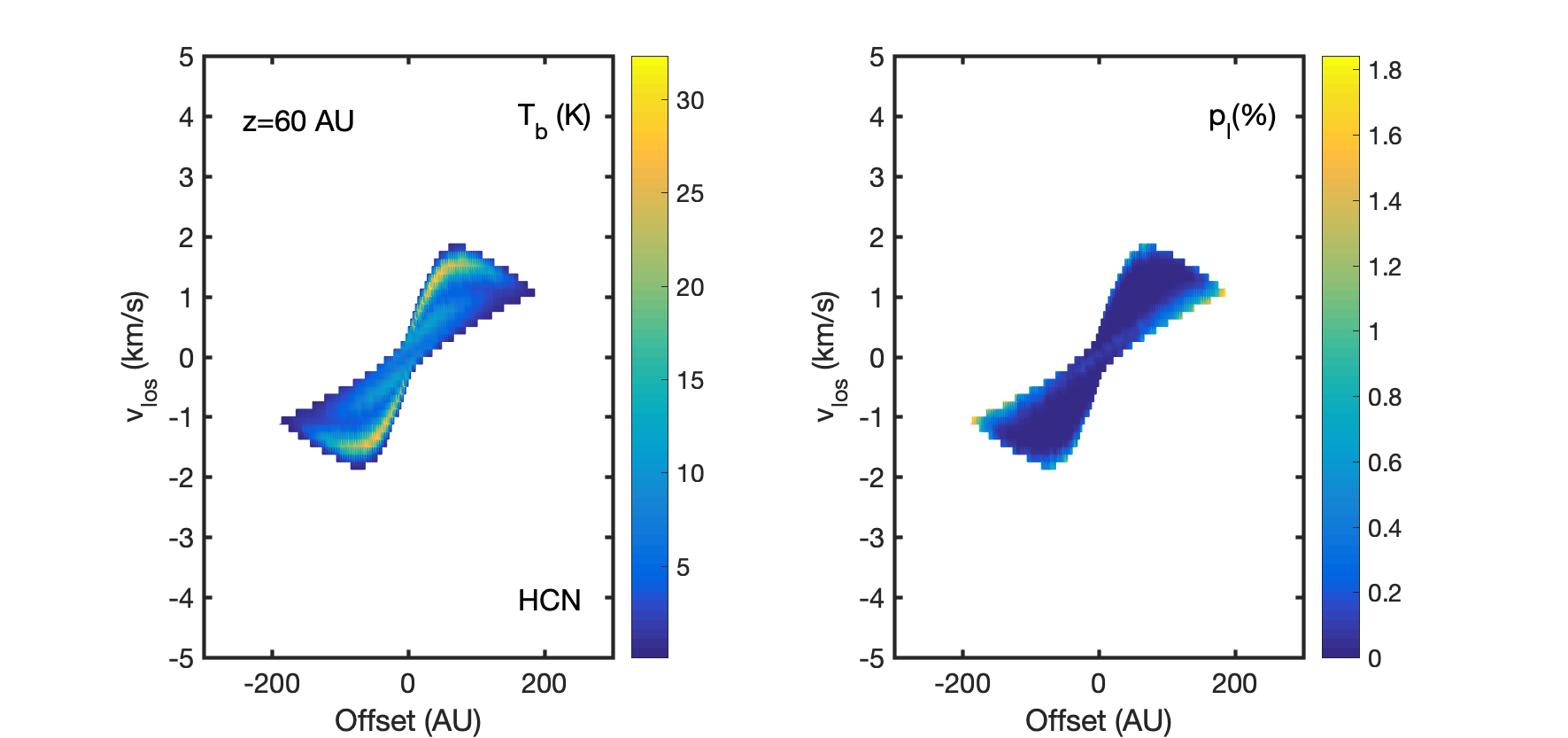}
    \caption{}
    \label{fig:45_pv1}
  \end{subfigure}
  \caption{Integrated intensity maps (a) and position-velocity diagrams (b) of the HCN $J=4\to 3$ transition. The PV diagram is taken at $z=60$ au. On the left-hand side, we plotted the total intensity, and on the right-hand side, the associated polarization fraction is shown. The magnetic field has a toroidal morphology.}
  \label{fig:45_both}
\end{figure*}

In Fig.~\ref{fig:45_HCN} we report the total integrated intensity and the polarization fraction at maximal emission of HCN emission of a $45^o$ inclined protoplanetary disk with a toroidal magnetic field configuration. We note that the predicted polarization is significantly weaker for this configuration in comparison to the edge-on oriented disk. Partially, this may be explained through the orientation of the magnetic field, which, at a $45^o$ inclined angle with respect to the line-of-sight, causes $\sin^2(45^o)=1/2$ times smaller polarization (see equation~(8) of \citet{lankhaar:20a}). Polarization fractions in excess of $0.2\%$ are only predicted in regions of weak emission. This behavior is also reflected in Fig.~\ref{fig:45_pv1}, where we report the PV diagram at height $z=60 \ \mathrm{au}$. Only in regions of weaker emission do we see significant polarization for a $45^o$ inclined disk.

\section{Discussion}
In the previous section, we focused on a fiducial disk model and estimates of the polarization fractions for CO and HCN. In this section, we explore the effects of variations of the fiducial disk model on the polarization estimates. We put our work in context by first discussing the relation of our 3D polarized radiative transfer calculations to traditional LVG GK modeling.
\subsection{Line polarization in protoplanetary disks}
The emergence of linear polarization in the line emission of nonparamagnetic molecules is a consequence of the partial alignment of the molecules. Molecules align themselves under the influence of an anisotropic radiation field because directional radiation has differential interaction probabilities for $\Delta m = \pm 1, 0$ transitions. In order for significant alignment to manifest, it is required that the rate of radiative interactions is comparable to, or larger than, collisional interactions. Additionally, it is required that the relative anisotropy of the radiation field \citep{lankhaar:20a}
\begin{align}
\delta_0^2 = \frac{\int d\Omega \ P_2 (\cos \theta) I_{\nu} (\Omega)/\sqrt{2}}{\int d\Omega \ I_{\nu} (\Omega)},
\end{align}
is significant ($\delta_0^2 > 1 \%$) at the frequencies resonant with transitions associated with the molecule under investigation. 

\begin{figure}[h!]
  \centering
  \begin{subfigure}[b]{0.45\textwidth}
    \includegraphics[width=\textwidth]{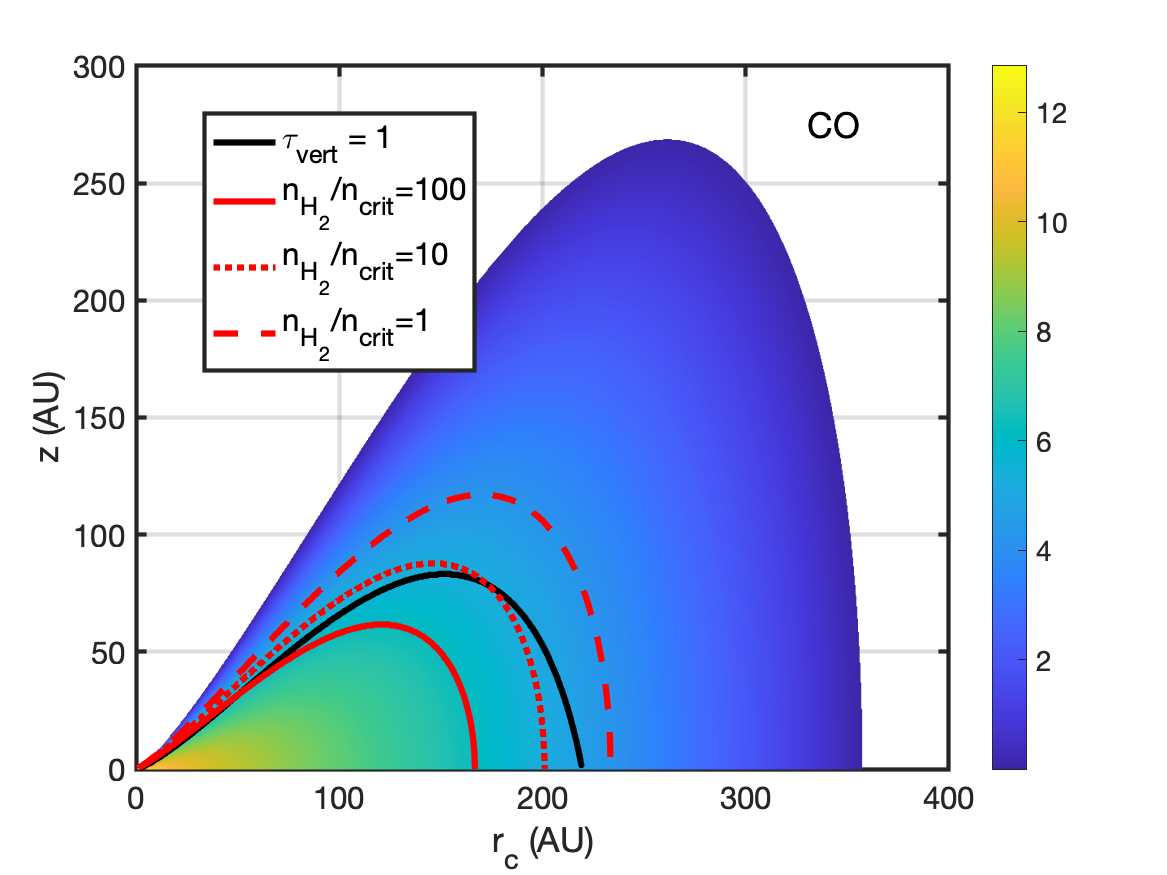}
    \caption{}
    \label{fig:AoC-tau_CO}
  \end{subfigure}
  ~
  \begin{subfigure}[b]{0.45\textwidth}
    \includegraphics[width=\textwidth]{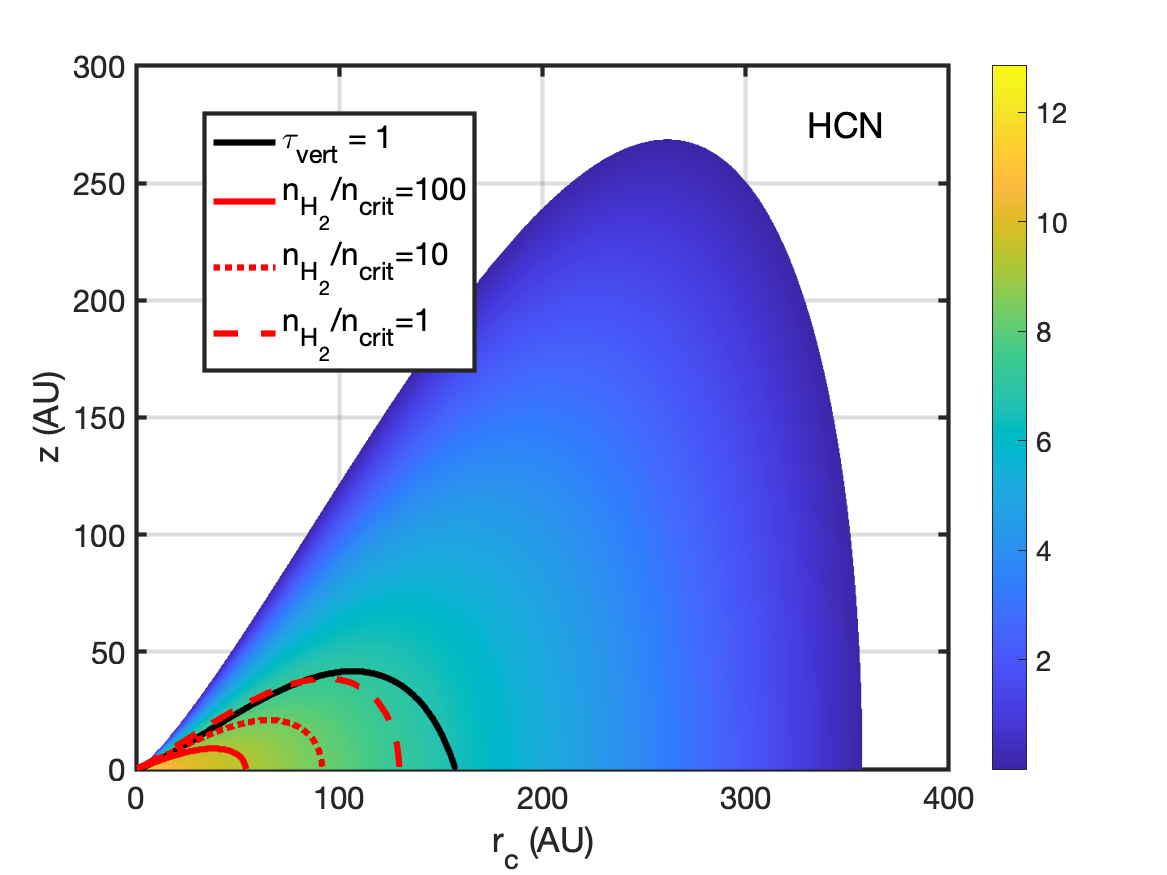}
    \caption{}
    \label{fig:AoC-tau_HCN}
  \end{subfigure}
  \caption{Density structure used in the protoplanetary disk polarization simulations, overlaid with the $\tau_{\mathrm{vert}}=1$ surface contour and the contours of critical density surfaces for CO (a) and HCN (b). The colorbar in the figure refers to the number density in units of cm$^{-3}$.}
  \label{fig:AoC-tau}
\end{figure}

An estimate of the ratio between radiative and collisional interactions for a given transition is obtained by comparing the disk density to the critical density. Table \ref{tab:AoC} reports the critical density of the transitions under investigation. In Fig.~\ref{fig:AoC-tau} we plotted the surface contours of the density in terms of the critical density for the CO $J=3\to2$ and HCN $J=4\to3$ transitions, using the fiducial disk model. 

It is more difficult to estimate the anisotropy of the radiation field. The classic result of \citet{goldreich:81} emphasized that maximal polarization emerges at optical depths $\sim 1$. In systems with $\tau \ll 1$, the anisotropy parameter $\delta_0^2$ of the radiation field is high, but radiative interactions are relatively weak, while at $\tau \gg 1$, radiative interactions are strong, but the radiation field isotropizes. In addition, in nonlocal simulations such as those with PORTAL, radiative anisotropy can also manifest when variations of the physical conditions, such as the temperature, over the optical depth are significant. 

In order to evaluate the applicability of the LVG GK formalism to a protoplanetary disk, we evaluated the general polarization characteristics by representing the excitation of molecular lines in the atmosphere of a protoplanetary disk as a plane-parallel slab problem. Along the $z$-axis of the disk, we assumed no velocity gradient, but in the $\hat{\boldsymbol{r}}_c\times \hat{\boldsymbol{z}}$ direction perpendicular to the cylindrical radius and the height, we have Keplerian rotation introducing a velocity gradient of $\frac{dv}{dr} = \Omega_{\mathrm{Kep}}$. 
Thus, we can get a measure for the optical depth in the disk atmosphere using the Sobolev approximation \citep{goldreich:81} 
\begin{align}
\tau_{\mathrm{LVG}} &= \frac{c (\kappa_{\nu} / \phi_{\nu})}{\nu_0 (dv/dr)}, 
\end{align}
where $\kappa_{\nu}$ and $\phi_{\nu}$ are the line opacity and profile. We evaluated the $\tau_{\mathrm{LVG}}$ at 1 scale height as a function of the cylindrical radius. We computed  $\tau_{\mathrm{LVG}}=1$ at $r_c \sim 150$ au for CO, while the other lines have $\tau_{\mathrm{LVG}}=1$ $\lesssim 100$ au. Comparing these results to the relative alignment of the molecular states, as presented in Fig.~\ref{fig:rel_all} and our results on the emergence of polarization in the emission, we note that in particular, regions with low LVG optical depth are significantly aligned and emit polarized emission. This is not expected from the theory of GK and points to a more complex polarization mechanism due to the complex (radiation) morphology of the protoplanetary disk. The complex radiation morphology of the protoplanetary disk arises because the gas properties change over a unit optical depth in large parts of the disks. The local approximation that is central to LVG prohibits one to take account of this. Additionally, in large parts of the disk atmosphere, the optical depth is $\lesssim 1$ in the vertical direction because of the density structure. This behavior is difficult to capture in the LVG approximation.

The vertical optical depth can provide us with an additional measure to estimate the polarization characteristics of molecular lines. In Fig.~\ref{fig:AoC-tau}, we plotted the $\tau_{\mathrm{vert}}=1$ surface of our protoplanetary disk model. Observing a disk face-on, radiation that has come from lower in the disk is almost completely reprocessed above the $\tau_{\mathrm{vert}}=1$ surface. Thus, it is the molecular alignment properties around the $\tau_{\mathrm{vert}}=1$ surface that determine the polarization characteristics of the emergent line emission.

It is striking to observe from Fig.~\ref{fig:AoC-tau} that the $\tau_{\mathrm{vert}}=1$ surface of both CO and HCN is located at densities low enough for significant radiative interactions to manifest. For CO, the $\tau_{\mathrm{vert}}=1$ surface is around $\frac{n_{\mathrm{H}_2}}{n_{\mathrm{crit}}} \sim 10$ for the $J=3\to 2$ transition, while for HCN, the $\tau_{\mathrm{vert}}=1$ surface is in a region of significant radiative interactions: $\frac{n_{\mathrm{H}_2}}{n_{\mathrm{crit}}} \lesssim  1$ for the $J=4 \to 3$ transition. Considering that radiative excitation is dominant in these regions, the polarization signal is limited by the radiation anisotropy in the region around the $\tau_{\mathrm{vert}}=1$ surface. 
 
\subsection{Polarization and disk characteristics}
\subsubsection{Disk mass}
\begin{figure}[h!]
  \centering
  \begin{subfigure}[b]{0.45\textwidth}
    \includegraphics[width=\textwidth]{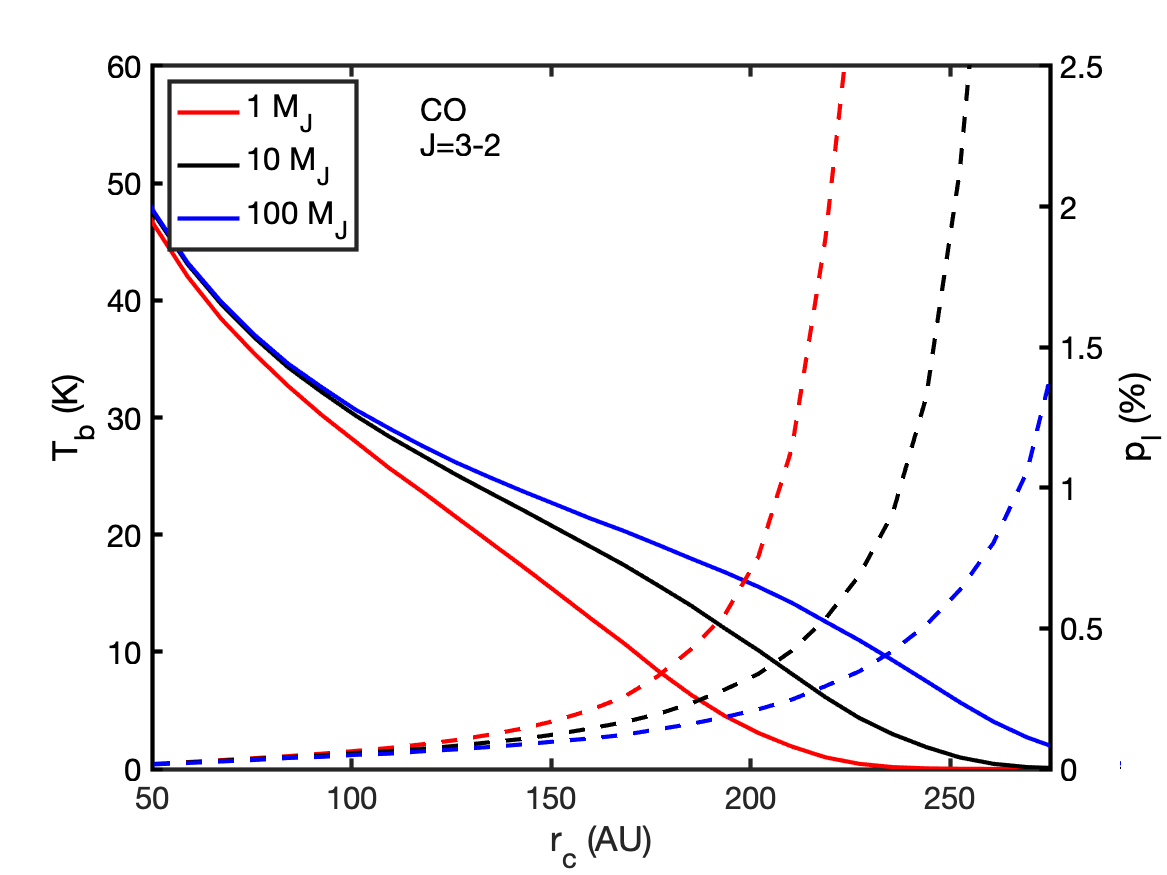}
    \caption{}
  \end{subfigure}
  ~
  \centering
  \begin{subfigure}[b]{0.45\textwidth}  
    \includegraphics[width=\textwidth]{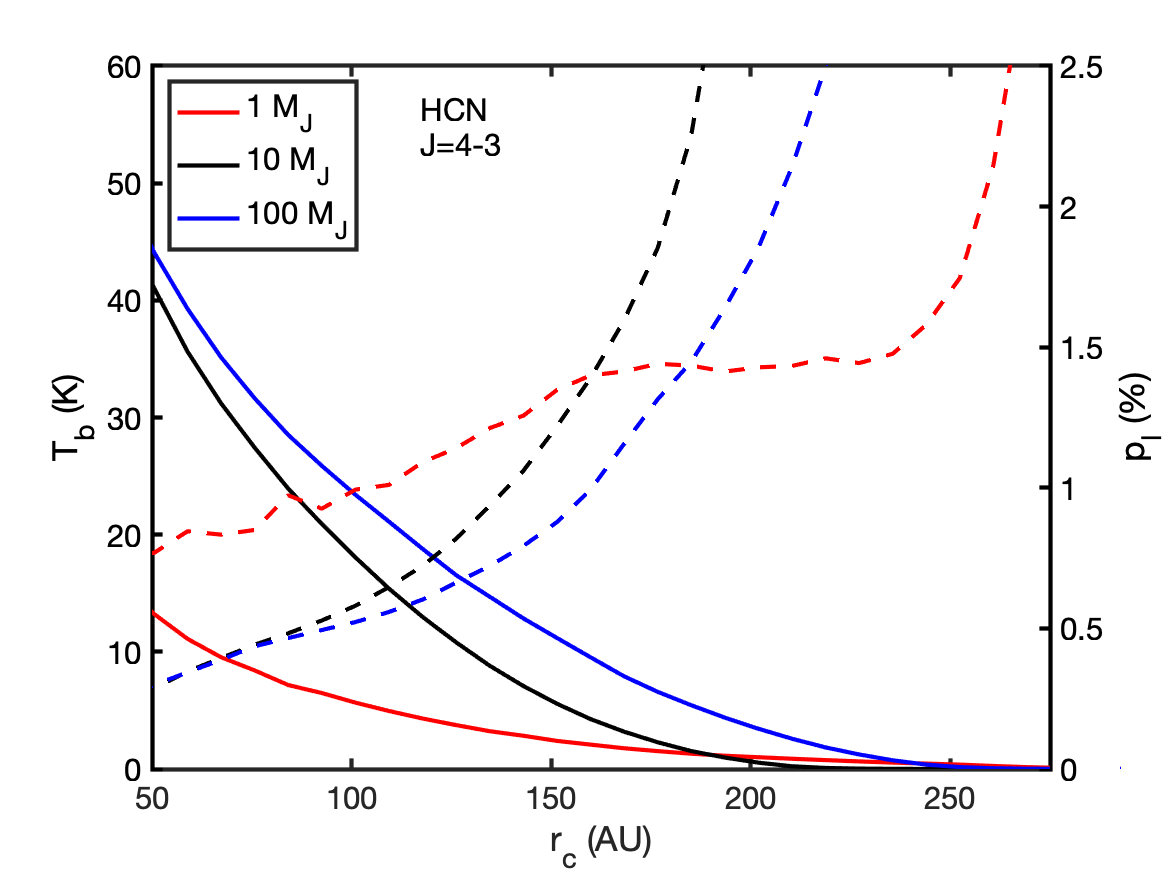}
    \caption{}
  \end{subfigure}
  \caption{Azimuthally averaged total emission and polarization fraction of transitions of CO (a) and HCN (b) from a face-on axisymmetric disk with variable mass. The polarization fraction was computed for a toroidal magnetic field configuration. The polarization fraction is omitted for $T_b < 0.1$ K.}
  \label{fig:spec_mass_range}
\end{figure}

In Fig.~\ref{fig:spec_mass_range}, we plotted polarization fraction estimates for the CO $J=3\to2$ and the HCN $J=4\to3$ lines for disk masses of $1$, $10$, and $100$ M$_{\mathrm{J}}$. The general trend of the polarization fraction is similar for all disk masses: we find the highest polarization fractions in the outskirts of the disk, where the density falls rapidly and emission quickly turns optically thin, while in the inner parts of the disk we predict lower polarization fraction. The polarization fraction estimates fall with increasing disk mass. The lower polarization fractions for more massive disks may be explained by their higher densities: keeping all other disk parameters constant, the density profile is proportional to the disk mass. This effect is mitigated by the $\tau=1$ surface lying higher up in the disk. Additionally, the total emission is higher for a more massive disk, so the emergent polarized flux density does not vary much with the disk mass.  

\subsubsection{Characteristic length}
\begin{figure}[h!]
  \centering
    \includegraphics[width=0.45\textwidth]{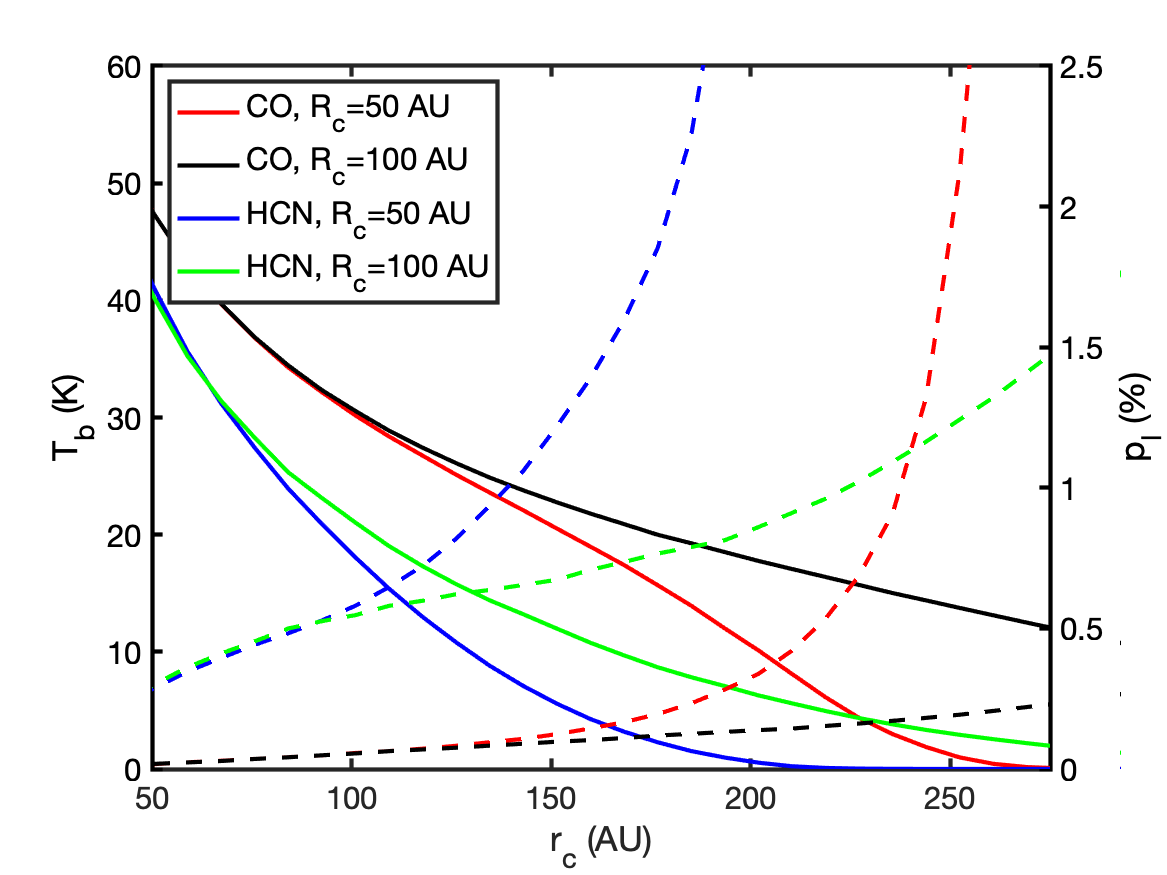}
  \caption{Azimuthally averaged total emission and polarization fraction of transitions of CO and HCN from a face-on axisymmetric disk with a variable characteristic length. The polarization fraction was computed for a toroidal magnetic field configuration. The polarization fraction is omitted for $T_b < 0.1$ K.}
  \label{fig:char_var}
\end{figure}

In Fig.~\ref{fig:char_var}, we present the polarization fraction of CO $J=3 \to 2$ and the HCN $J=4\to3$ transitions for face-on disks with characteristic lengths of $R_c=50$ and $100$ au. We note that at low projected distances to the star, the polarization fractions and total emission of both disks are rather similar, with the polarization estimates being slightly higher for larger characteristic length. At higher characteristic lengths, the tapering of the disk sets in farther away from the central (proto)star. The disk mass is therefore more smeared out for larger characteristic lengths, leading to slightly lower densities close to the protostar. At a larger distance from the star, we note that that the emission is sustained farther out for the $R_c=100$ au simulations, as well as that the polarization is consistently weaker in the disk with a larger characteristic length. We may ascribe this to higher densities farther out in the disk, leading to both higher collisional rates, but also to a weaker intensity gradient, and thus lower radiation anisotropy. 

It should be added to this analysis that we used a constant abundance profile for our chemical species. In particular, freeze out processes occur in the colder regions of the disk at a large distance from the central stellar object. An abundance drop in these regions mimics the tapering that would otherwise come from a disk with a low characteristic length. So for disk models that would include chemical modeling and a realistic abundance profile, we do not expect much variation in the polarization properties of molecular spectral lines with the characteristic length of the disk model. 
  
\subsubsection{Vertical profile}
We have represented the vertical profile of a disk by by assuming a Gaussian distribution across the midplane $\rho (z) \propto \exp(-(z/H)^2)$, where the scale height, $H$, is a function of the cylindrical radius as in Eqs.~(\ref{eq:density}). However, such a profile corresponds to an isothermal (in the $z$-direction) disk, and it does not include a realistic magnetic pressure. At larger $z/H$, such a profile underestimates the density, while it is precisely in these regions where we expect our molecular species to be excited. The high concentration of matter in the central parts of the disk also leads to a $\tau_{\mathrm{vert}}$ surface that is closer to the midplane. For our purposes of simulating the emergence of polarization, it is important that we check the robustness of our results with respect to the vertical gas-density distribution. A vertical profile that includes the contribution of the magnetic pressure can be approximated by \citep{armitage:19, turner:14, hirose:11}
\begin{align}
\rho (r_c,z) = \frac{\Sigma(r_c)}{\sqrt{2 \pi}h(1+\epsilon)} \left[e^{-(z/\sqrt{2}H)^2} + \frac{\sqrt{4\pi} \epsilon}{3} e^{-2|z|/3H} \right],
\end{align}
and it has an additional non-Gaussian profile that dominates for larger $z/H$. We performed simulations with such an extended profile and indeed observed slightly larger ($\sim 10\%$) polarization fractions for all molecules tested. This small difference lends confidence to approximating the vertical density profile with a Gaussian distribution for our purposes. The difference may be ascribed to a higher $\tau_{\mathrm{vert}}$ surface, where anisotropy of a local physical structure manifested more strongly in the radiation anisotropy. 

\subsubsection{Substructures}
Recent observations reveal pronounced substructures in a variety of protoplanetary disks in both the dust and molecular spectral emission \citep{andrews:18, huang:18}. These substructures may be concentric or local, and they can be associated with intensity depressions or elevation, linked to pressure, density, and temperature gradients. In this work, we have not investigated the possible influence of substructures on the polarization profile of protoplanetary disks. However, considering that the presence of radiation anisotropy is the limiting factor in the emergence of line polarization, and that substructures lead to local radiation anisotropy, it may be hypothesized that substructures will be associated with high levels of line polarization. Polarization observations would thus provide an interesting opportunity to investigate the nature of these substructures, particularly their connection to the magnetic field \citep{suriano:18}. We leave such investigations for future work.

\subsection{Polarization and line characteristics}
\subsubsection{Molecular transition}
\begin{figure}[h!]
  \centering
  \begin{subfigure}[b]{0.45\textwidth}
    \includegraphics[width=\textwidth]{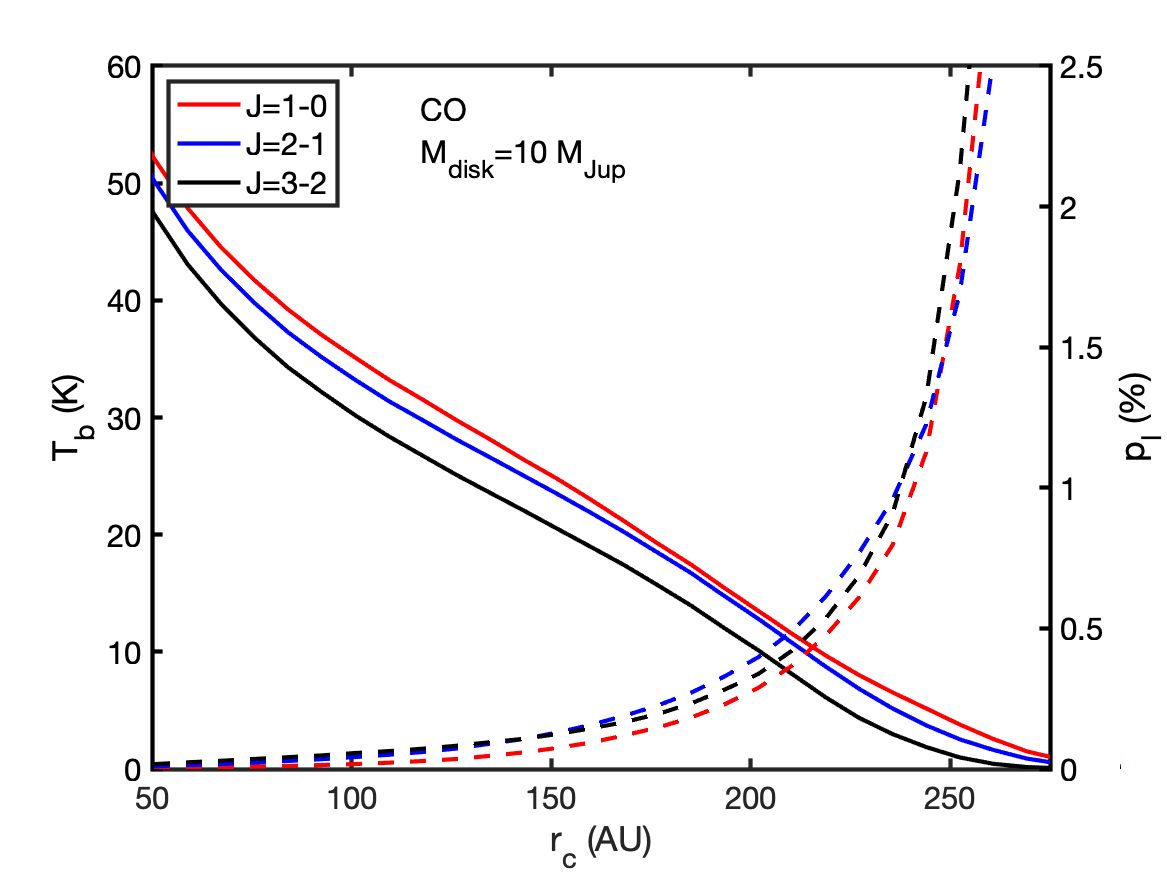}
    \caption{}
  \end{subfigure}
  ~
  \begin{subfigure}[b]{0.45\textwidth}
    \includegraphics[width=\textwidth]{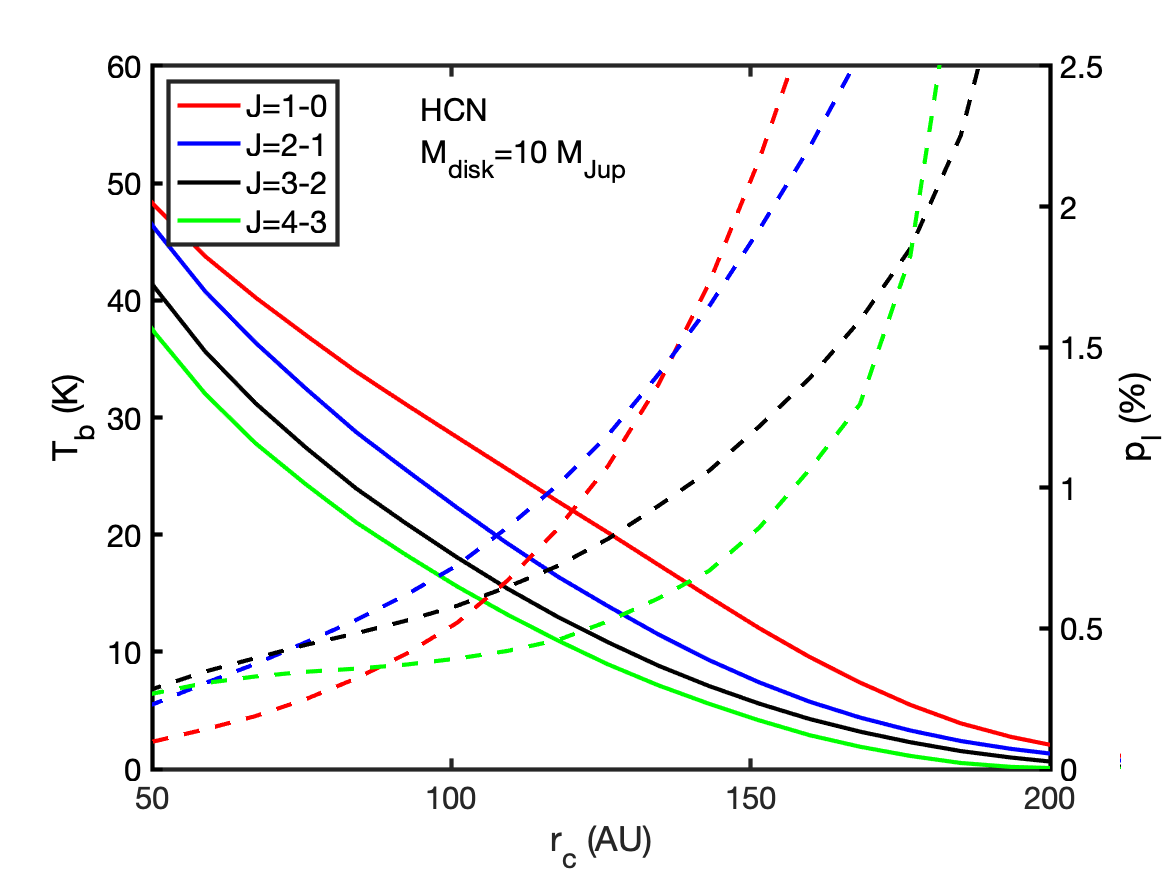}
    \caption{}
  \end{subfigure}
  \caption{Azimuthally averaged total emission and polarization fraction of a variety of transitions of CO (a) and HCN (b) from a face-on axisymmetric disk with a mass of 10 $M_{\mathrm{J}}$. The polarization fraction was computed for a toroidal magnetic field configuration. The polarization fraction is omitted for $T_b < 0.1$ K.}
  \label{fig:spec_line_range}
\end{figure}

In the simulations we have presented thus far, we used molecular lines that are in ALMA's band 7. In Fig.~\ref{fig:spec_line_range}, we report the polarization properties of other transitions of HCN and CO. The general trend that we observed earlier, where lines are polarized most strongly in the outer part of the disk while polarization fractions drop toward the protostar, is seen in all transitions of both HCN and CO. In fact, the polarization properties of the different lines of CO are very similar between the $J=1\to0$, $J=2\to1$, and $J=3\to2$ transitions. We may explain this behavior through a combination of circumstances that affect the polarization of these lines. The same level of alignment produces higher polarization fractions for lines with low angular momentum \citep{lankhaar:20b}, but, on the other hand, collisional interactions become increasingly important for lower levels, thus lowering the levels of alignment.

The polarization of the different HCN lines shows more complicated behavior. In the inner parts of the disk, for $r_c<100$ au, the $J=2\to1$, $J=3\to2$, and the $J=4\to3$ transitions show a similar polarization behavior, but we predict significantly lower polarization for the $J=1\to0$ line. At large projected distances from the protostar, the predicted polarization fraction of the $J=1\to0$ transitions rises quickly. We observe that the steep rise in the polarization fraction occurs at a larger projected distance from the central star for the higher frequency transitions. 

The different transitions of the same species trace different regions of the protoplanetary disk and thus may theoretically be used to probe the magnetic field properties at different regions of the disk. From our predictions, however, the question of whether all transitions give rise to detectable polarization flux densities is raised. Because the polarization fraction properties are similar for all transitions, this means that polarized flux densities of lower frequency lines are comparatively small. Later on in this section, we discuss the feasibility of detecting polarization in these lines. 

\subsubsection{Isotopologues of CO}
\begin{figure}[h!]
  \centering
  \includegraphics[width=0.45\textwidth]{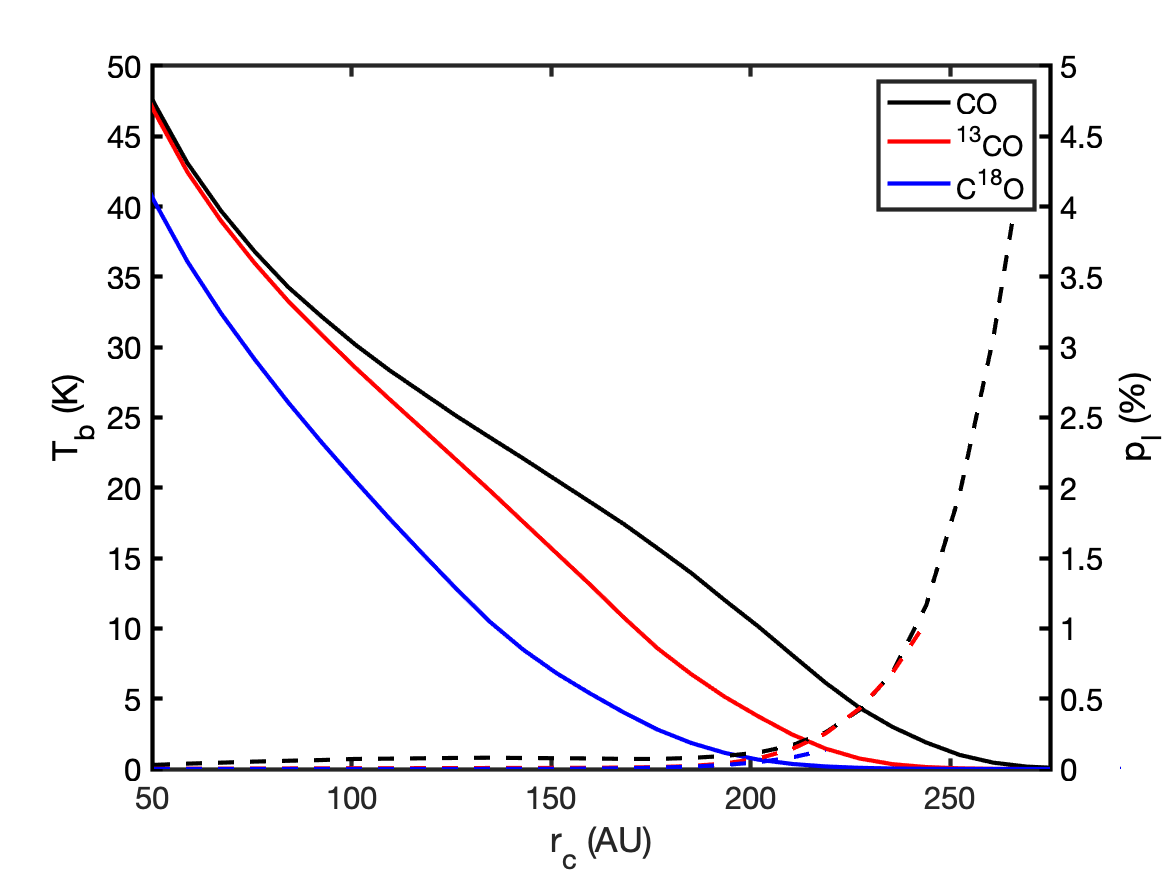}
  \caption{Azimuthally averaged total emission and polarization fraction of transitions of isotopologues of CO from a face-on axisymmetric disk. The polarization fraction was computed for a toroidal magnetic field configuration. The polarization fraction is omitted for $T_b < 0.1$ K.}
  \label{fig:spec_isot_tor}
\end{figure}

The radiative and collisional rates between the CO isotopologues are comparable at any point in the disk. However, the emission emerges from closer toward the midplane for the less abundant $^{13}$CO and C$^{18}$O, compared to the main isotopologue of CO \citep[see also,][]{zhang:17}. In other words, the $\tau_{\mathrm{vert}}=1$ surface of $^{13}$CO and C$^{18}$O is closer to the midplane compared to the $\tau_{\mathrm{vert}}=1$ surface of $^{12}$CO. Thus, $^{12}$CO traces more diffuse gas than its isotopologues. Especially in the dense environments that characterize the protoplanetary disk, this means that $^{12}$CO is consistently more polarized than its isotopologues in the optically thick parts of the disk, as may be seen in Fig.~\ref{fig:spec_isot_tor}. Also, in optically thin parts of the line emission, $^{12}$CO is more polarized because its significant emission extends farther out than $^{13}$CO and C$^{18}$O, and thus it traces more diffuse gas. 

A similar effect can be observed when one analyzes the spectrum of the polarization of $^{12}$CO emission lines. For a face-on disk, maximum polarization is found at the line center, where the emission originates farthest from the midplane, whereas off-resonance, radiation comes from closer to the disk midplane. We may therefore expect maximal polarization of lines at their line center, at least in the simple case of the emission coming from a face-on disk. 

\subsubsection{Effects of vibration}
In this work, we have considered molecules that have close-lying states that may be radiatively excited through FIR radiation. As an example, if we focus on the triatomic molecules, HCN, HCO$^+$, N$_2$H$^+$, and C$_2$H, they have bending modes with associated wave numbers of $\tilde{\nu}=760 \ \mathrm{cm}^{-1}$, $829 \ \mathrm{cm}^{-1}$, $687 \ \mathrm{cm}^{-1}$, and $372 \ \mathrm{cm}^{-1}$, respectively, while the Einstein coefficients of vibrational transitions are in the order of $\sim \mathrm{s}^{-1}$. Consequently, thermal radiation of $\sim 150 \ \mathrm{K}$ interacts with levels of these molecules with similar interaction rates as collisions would around $n_{\mathrm{H}_2} \sim 10^8 \ \mathrm{cm}^{-3}$. In practice, this means that at the denser and warmer inner regions, it may be hypothesized that alignment manifests through these vibrational transitions. 

In addition to an enhancement of the polarization fraction, strong radiative interactions with higher vibrational states may alter the symmetry axis of the molecule. In PORTAL, we assume that the symmetry axis of the molecules is along the magnetic field direction. This is the case when the magnetic precession rate ($g \Omega \sim \mathrm{s}^{-1}/\mathrm{mG}$) is $10-100$ times higher than any other aligning interaction. Interactions with a low-lying vibrational state via a strong radiation field may occur at similar rates as the magnetic precession rate. This gives rise to the Hanle effect, and it may be associated with an increase in the polarization fraction and the rotation of the molecular alignment direction, thus also leading to a rotation in the polarization angle of the emission \citep{landi:06, lankhaar:19}. 
 
We performed simulations including the low-lying bending modes ($760 \ \mathrm{cm}^{-1}$) of HCN. We assumed vibration level-changing collisional transitions to be negligible with respect to radiative vibrational transitions \citep{ziurys:86}. We computed the vibrational transition Einstein coefficients using equation~(13) of \citet{ramos:05} with $A_0 = 3.7 \ \mathrm{s}^{-1}$ \citep{ziurys:86}. We found that the estimated polarization fractions of transitions in the ground state are unaffected by including radiative interactions with the vibrational bending mode. The radiation field around $760 \ \mathrm{cm}^{-1}$ is too weak to interact with ground state levels of HCN at a high rate. We found that a small fraction of the HCN may be found in the vibrationally excited bending mode, and that these molecules are significantly aligned, much more so than in the vibrational ground state. These results lead to the conclusion that vibrational interactions may be ignored when analyzing the alignment of molecules in protoplanetary disks. Still, at other stages of star formation or in other sources, effects of vibrational transitions might be of importance and may enhance polarization fractions significantly. 
 
Other molecules that are sensitive to FIR radiation are rotors with relatively light constituents such as NH$_3$ or H$_2$O. Radiative interactions of these molecules happen at rather high rates because of the large rotational energy these molecules posses per quantum. Critical densities of these molecules are accordingly anomalously high \citep{elitzur:92}. Alignment can already manifest itself at high densities for these types of molecules and they might be of interest for future polarization studies. 

\subsection{Observational feasibility of spectral line polarization}
\begin{figure}[h!]
  \centering
  \includegraphics[width=0.45\textwidth]{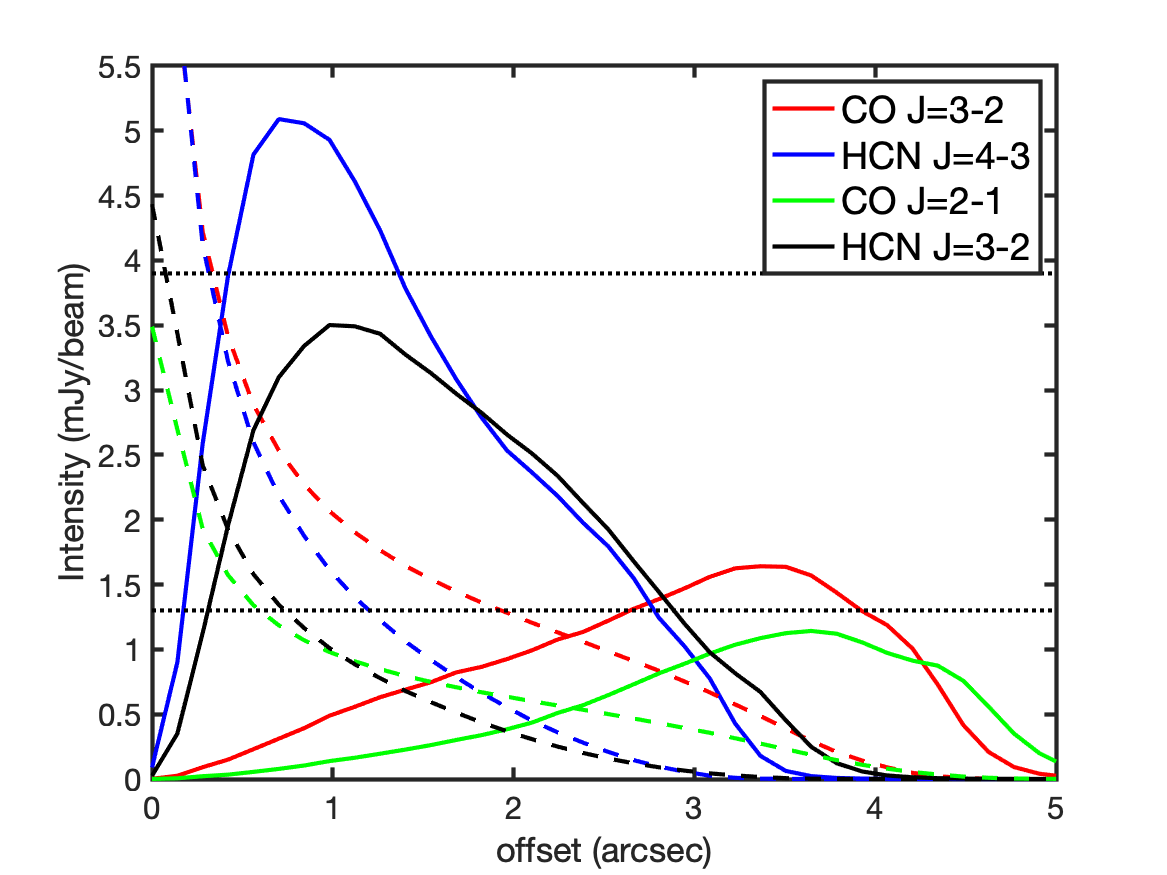}
  \caption{Azimuthally averaged polarized emission of transitions of HCN and CO from a face-on axisymmetric disk of mass $10 \ M_{\mathrm{J}}$. We plotted the $J=3 \to 2$ and $J=2 \to 1$ transitions of CO and the $J=4\to 3$ and $J=3\to 2$ transitions of HCN. We assumed a source of distance $60$ pc and a beam size (FWHM) of $0.7 \arcsec \times 0.7\arcsec$. The polarized emission was computed for a toroidal magnetic field configuration. We also plotted the total emission divided by $1000$ associated with each transition in a dotted line of the same color to represent the (current) ALMA sensitivity limit of the detectability of linear polarization. We also added a dotted line, representing $1 \times$ and $3\times$ a typical ALMA sensitivity of $1.3$ mJy.}
  \label{fig:TW_Hya}
\end{figure}

The GK effect has yet to be detected in protoplanetary disks. Recently, \citet{stephens:20} report upper limits of $3\%$ on the CO (and isotopologues) polarization fraction in the inclined protoplanetary disk HD 142527 and IM Lupus. Indeed, our simulations find low polarization fractions for CO in the inner regions of the disk, while in the outer regions, polarization fractions are high, but the polarized flux density remains under the detectability threshold (see Fig.~\ref{fig:TW_Hya}). \citet{stephens:20} speculate that the isotopologues of CO are better candidates for polarization observations because their optical depths are around unity in disk regions of interest, but we find the opposite to be the case. While it is the case that CO isotopologue optical depths are lower, and closer to unity in the inner parts of the disk, as a consequence, this emission comes from deeper in the disk, where densities are higher and tend to thermalize the energy levels. Thus the alignment necessary for the production of polarization for these species is effectively quenched. We note that $^{12}$CO emission comes from higher in the disk atmosphere, but, indeed, its high optical depth tends to isotropize the radiation close to the protostar, thus also suppressing the polarization fraction.

For disks $\gtrsim 10 \ M_{\mathrm{J}}$, CO is only marginally aligned in the regions where they emit strongly. Only in the diffuse outer parts of the protoplanetary disk do they emit radiation that is significantly polarized. Instead of CO observations, observers may be advised to use molecules with higher dipole moments, such as HCN, as a magnetic field tracer in protoplanetary disks. In our simulations, we have found that HCN is aligned in a larger part of the disk, both through its lower optical depth, and because of its higher critical density. We predict higher polarization fraction estimates for this species for any inclination. The polarized emission intensity from HCN is highest close to the central protostar. We have not found a strong dependence of the polarization fraction on the transition, only the tendency that low $J$ transitions are less polarized in the inner parts, but higher in the outer parts. Low mass disks are predicted to lead to high polarization fractions. However, the polarization quenching effect that one might expect for higher mass disks is not so pronounced because the emission from more massive disks comes from higher up in the disk atmosphere.

Other molecules that have high dipole moments are the molecular ions HCO$^+$ and N$_2$H$^+$. Estimated polarization fractions of these species are lower than HCN because the collisional cross sections between ions and H$_2$ are relatively large. However, in these PORTAL simulations, we have not included the possible alignment-enhancing directional collision interactions, which could be instantiated through ambipolar diffusion \citep{lankhaar:20b}. Particularly toward the midplane of the disk, ionization fractions are low, and thus would predict high ambipolar drift velocities \citep{cleeves:15, mouschovias:99}. 

We have presented polarization fraction estimates for a face-on, edge-on, and $45^o$-inclined disks. In the appendix, we present polarization maps overlaying the polarization vectors over total intensity maps. A face-on disk presents us with the simplest geometry and the interpretation of the polarization results is the most straight forward. Face-on disk polarization observations are not sensitive to a poloidal magnetic field configuration. From the polarization vector maps (Appendix), we note that for any inclination, the polarization vectors from toroidal or poloidal magnetic field geometries are oriented parallel to the magnetic field direction. While for a radial magnetic field, the polarization vectors are oriented perpendicular to the magnetic field direction. A toroidal and radial magnetic field is readily distinguished from a poloidal magnetic field through the polarization morphology. A toroidal magnetic field may be distinguished from a radial magnetic field for inclined disks through the dependence of the polarization morphology on the velocity channel.   

The polarization fractions we report might be underestimated due to the anisotropic intensity approximation \citep{lankhaar:20a}. In PORTAL, we only considered the total intensity in formulating the aligning part of the radiation interaction. In \citet{lankhaar:20a}, it is shown that this leads to an underestimation of about half a percent in regions of high optical depth and low critical densities.

Finally, we estimated the observational viability of HCN and CO polarization measurements of the well-known protoplanetary disk around TW Hya. TW Hya is almost face-on, with an inclination of $i = 5^o$. It is at a distance of $60$ pc and has a disk mass of about $10 \ M_{\mathrm{J}}$ \citep{huang:18}. In Fig.~\ref{fig:TW_Hya}, we plotted estimates of the polarized emission of the $J=3 \to 2$ and $J=2 \to 1$ transitions of CO and the $J=4\to 3$ and $J=3\to 2$ transitions of HCN, while using a beam size (FWHM) of $0.7 \arcsec \times 0.7\arcsec$. We note that the current ALMA detection limit is $0.1\%$ for linear polarization. It is apparent that CO emission has its significant polarization in the outer parts of the disk, while HCN has its most strongly polarized emission around $1 \arcsec$ (60 au). We note here that higher frequency transitions have their peak polarized emission slightly closer to the disk center. The polarized emission is much stronger for HCN than for CO, and it is doubtful that polarized emission from CO is detectable for a $10 \ M_{\mathrm{J}}$ disk. According to our predictions, HCN emission should have a detectable polarization between $1 \arcsec$ and $3 \arcsec$. 

\section{Conclusions}
We present polarization estimates of molecular spectral lines that are excited in protoplanetary disk regions. We report on the polarization characteristics of different molecules and conclude that those with strong dipole moments and relatively low collision rates are the most useful for polarization measurements. We consider HCN to be typical of such a species, and we found its spectral lines to be significantly polarized in large regions of the protoplanetary disk. The molecule CO, which has been traditionally observed as a molecule exhibiting polarization in its line emission, is less suitable for polarization observations in disks. Even though the abundance of CO is high, and its visible emission comes from high up in the disk, where the gas is diffuse, its high optical depth isotropizes the radiation, thus leading to marginal polarization in its line emission. We note that CO isotopologues that exhibit a lower optical depth are excited in the lower regions of the disk, where collisions dominate the excitation and thus no alignment can manifest itself, leading to even lower polarization estimates for these species compared to regular $^{12}$CO. 

We investigated the influence of a number of disk properties on the polarization properties of molecular lines. We found that the disk characteristic that impacts polarization characteristics of molecular lines most strongly is the disk mass. Still, for lines that are optically thick, the influence of the disk mass is mitigated through these lines emerging from higher up in the disk, where the gas is more diffuse. We predict that low mass disks exhibit high line polarization fractions in the spectral lines, but significant polarization in the line emission of high mass disks is also present. 

Our simulations indicate that molecules with strong dipole moments and relatively weak collisional rate coefficients, such as HCN, should be targeted for polarization observations. Such species will reveal the magnetic field configuration of the inner parts of the disk, and they may yield important information about the accretion dynamics of protoplanetary disks. 

\begin{acknowledgements} 
Support for this work was provided by the Swedish Research Council (VR) through grant No. 2014-05713. 
\end{acknowledgements}

\bibliography{/Users/boylankhaar/texlibs/lib}

\begin{appendix}
\section{Additional figures}
\begin{figure*}
    \centering
    \begin{subfigure}[b]{0.32\textwidth}
       \includegraphics[width=\textwidth]{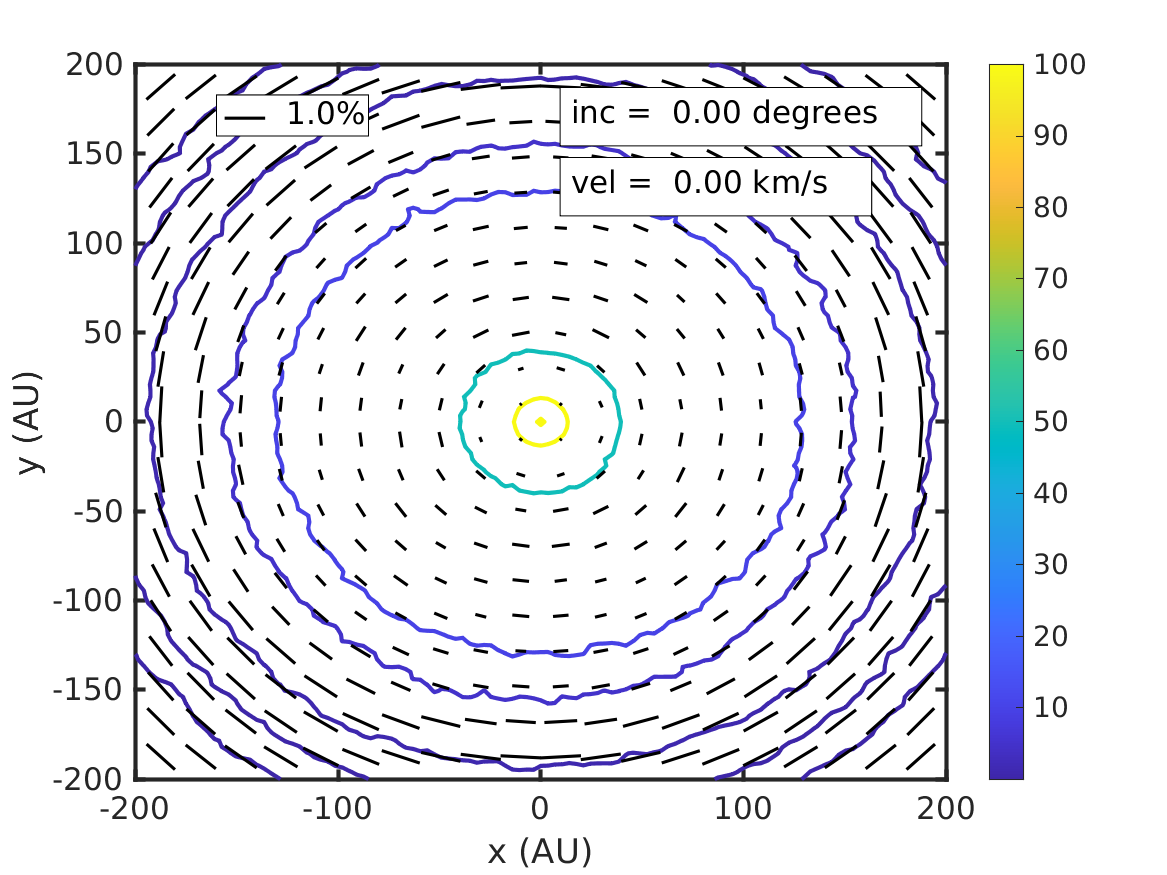}
       \caption{}
    \end{subfigure}
    ~
    \begin{subfigure}[b]{0.32\textwidth}
       \includegraphics[width=\textwidth]{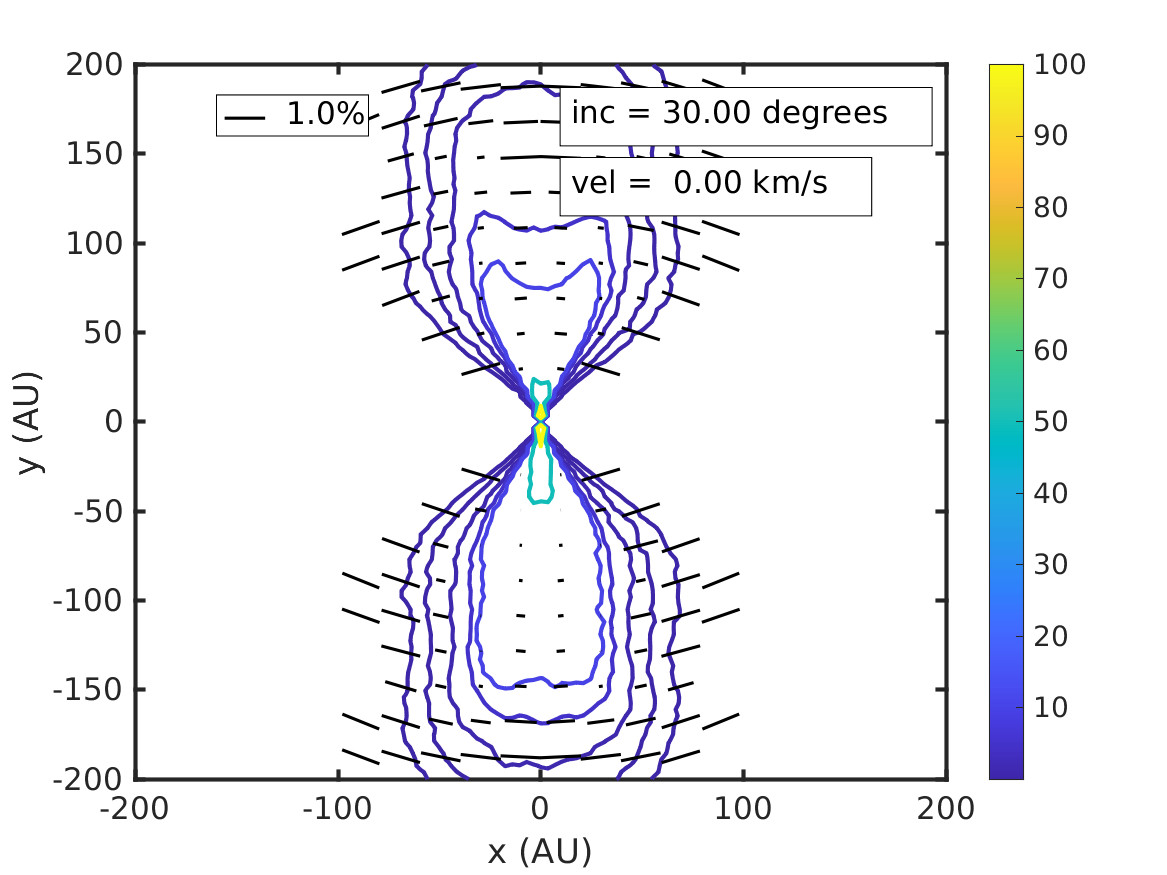}
       \caption{}
    \end{subfigure}
     ~
    \begin{subfigure}[b]{0.32\textwidth}
       \includegraphics[width=\textwidth]{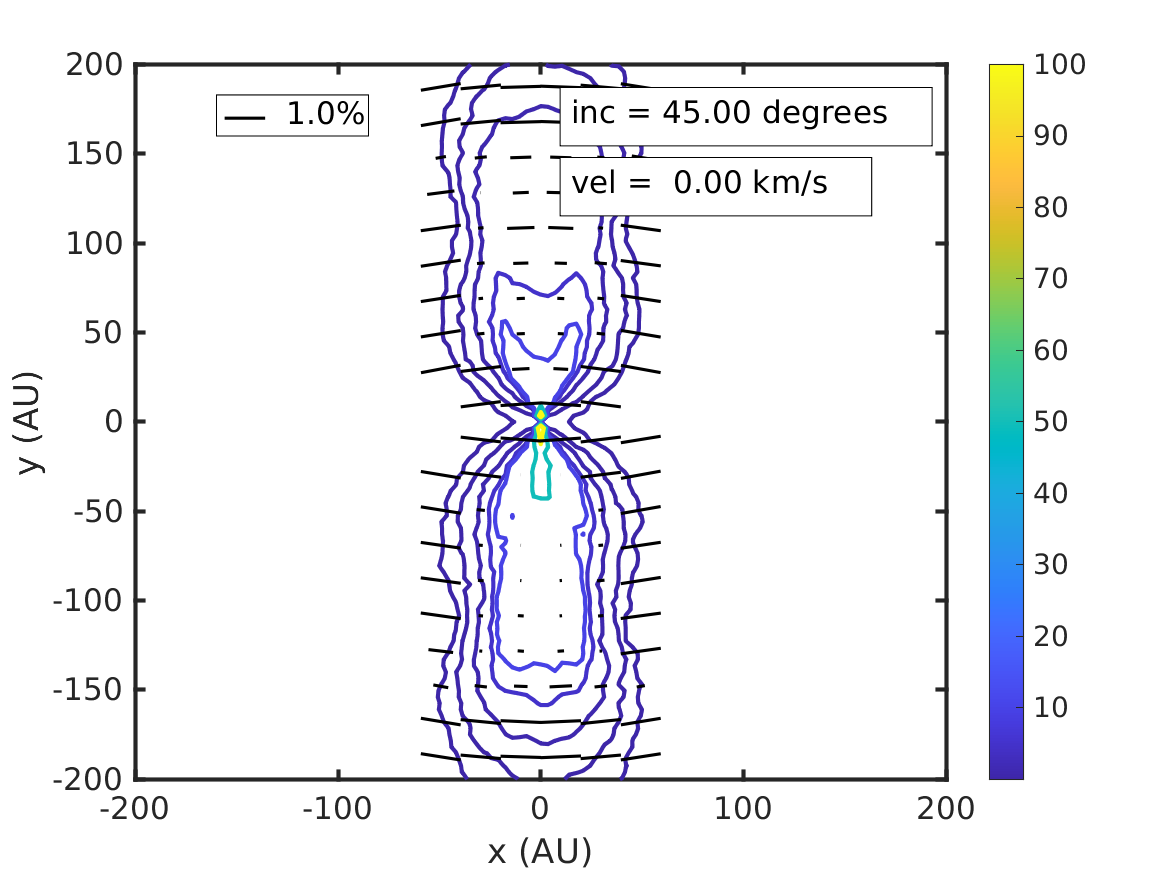}
      \caption{}
    \end{subfigure}
     ~
    \begin{subfigure}[b]{0.32\textwidth}
       \includegraphics[width=\textwidth]{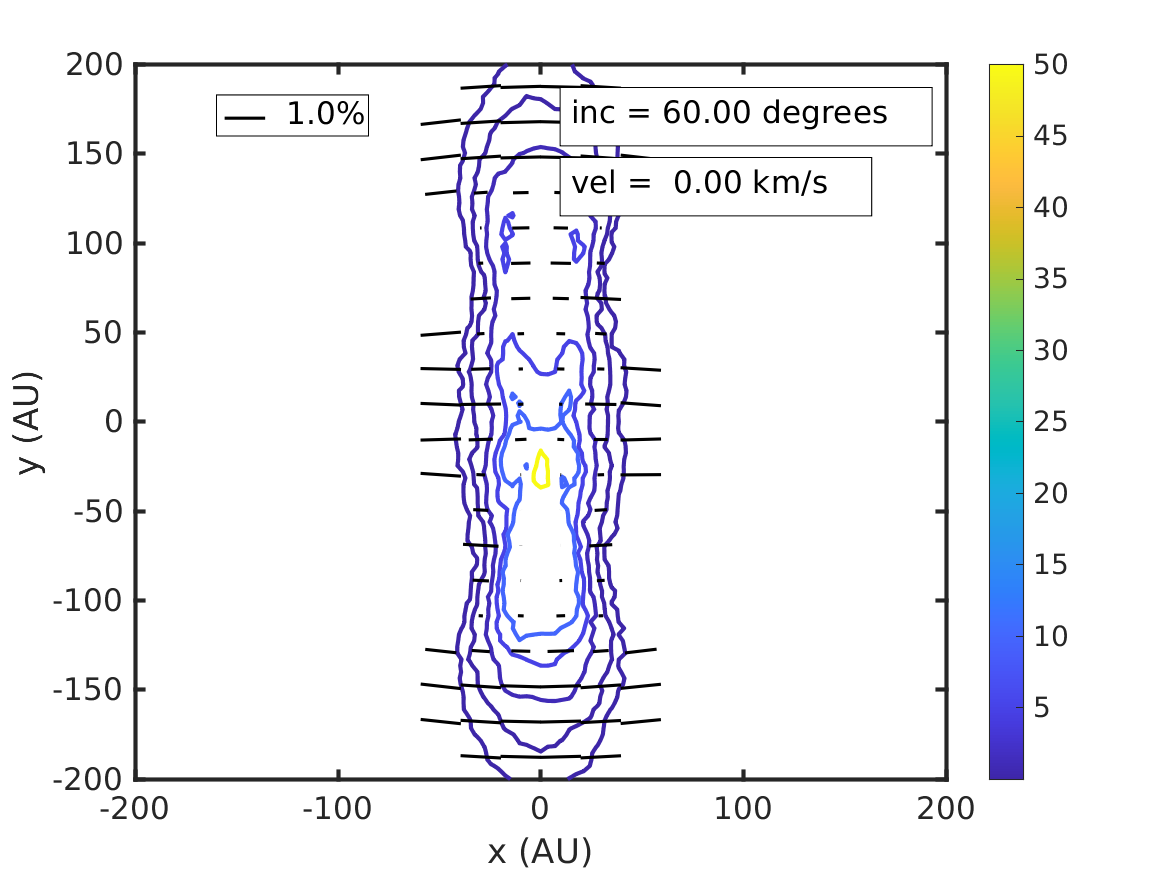}
       \caption{}
    \end{subfigure}
    ~
    \begin{subfigure}[b]{0.32\textwidth}
       \includegraphics[width=\textwidth]{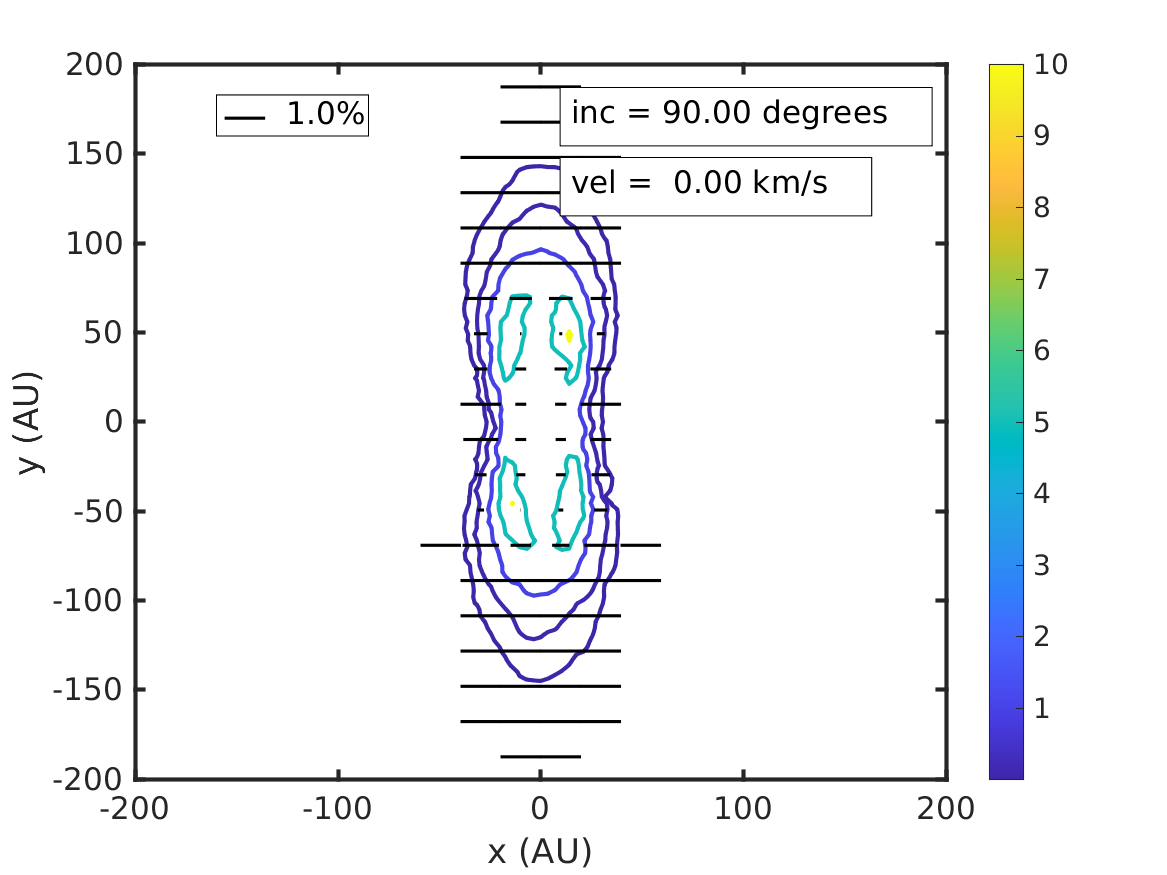}
       \caption{}
    \end{subfigure}

    \begin{subfigure}[b]{0.32\textwidth}
       \includegraphics[width=\textwidth]{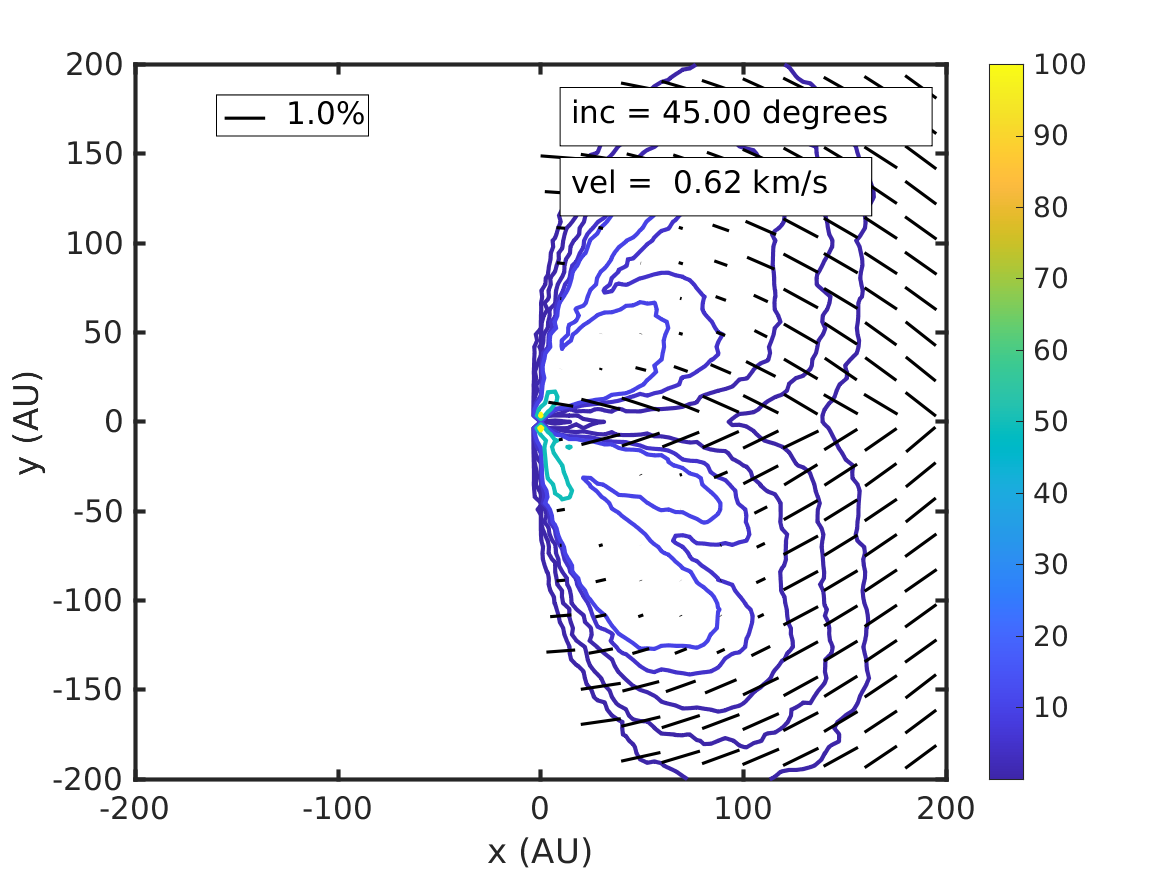}
      \caption{}
    \end{subfigure}
     ~
    \begin{subfigure}[b]{0.32\textwidth}
       \includegraphics[width=\textwidth]{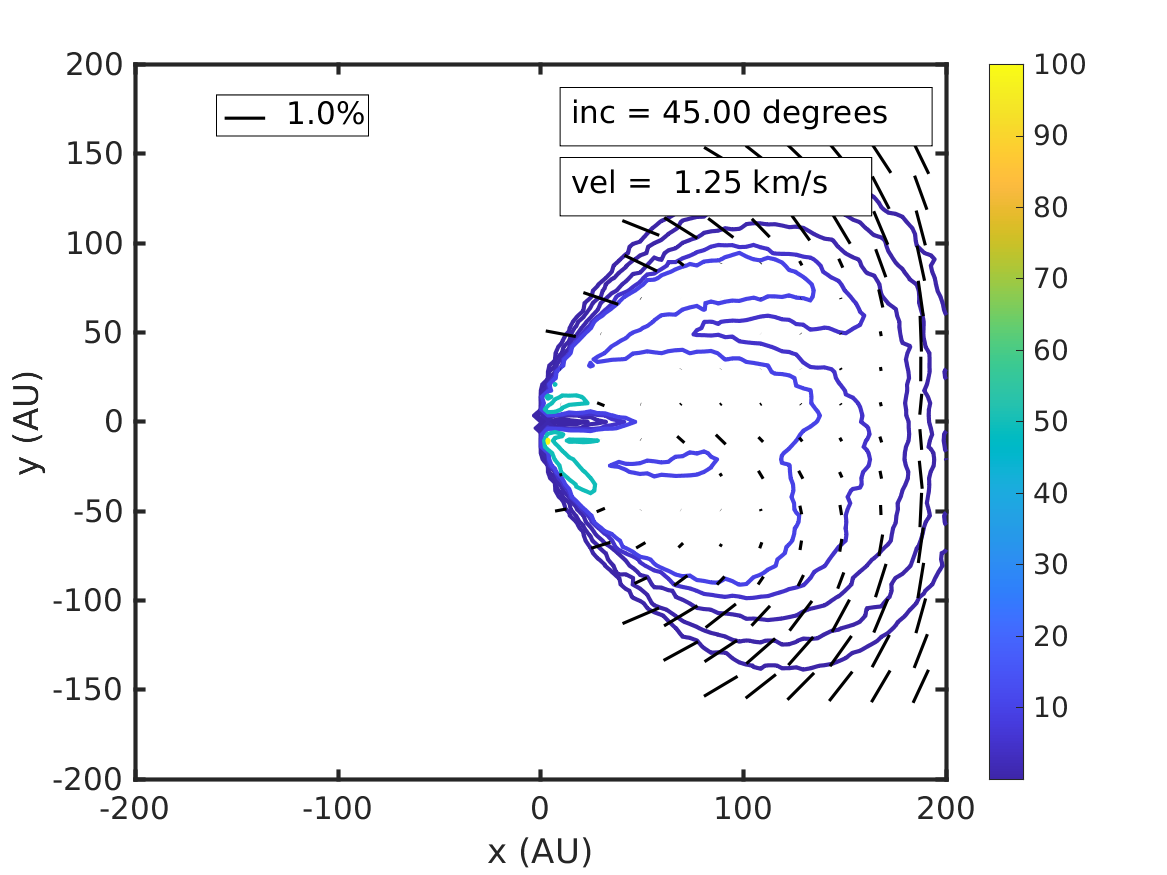}
       \caption{}
    \end{subfigure}
    ~
    \begin{subfigure}[b]{0.32\textwidth}
       \includegraphics[width=\textwidth]{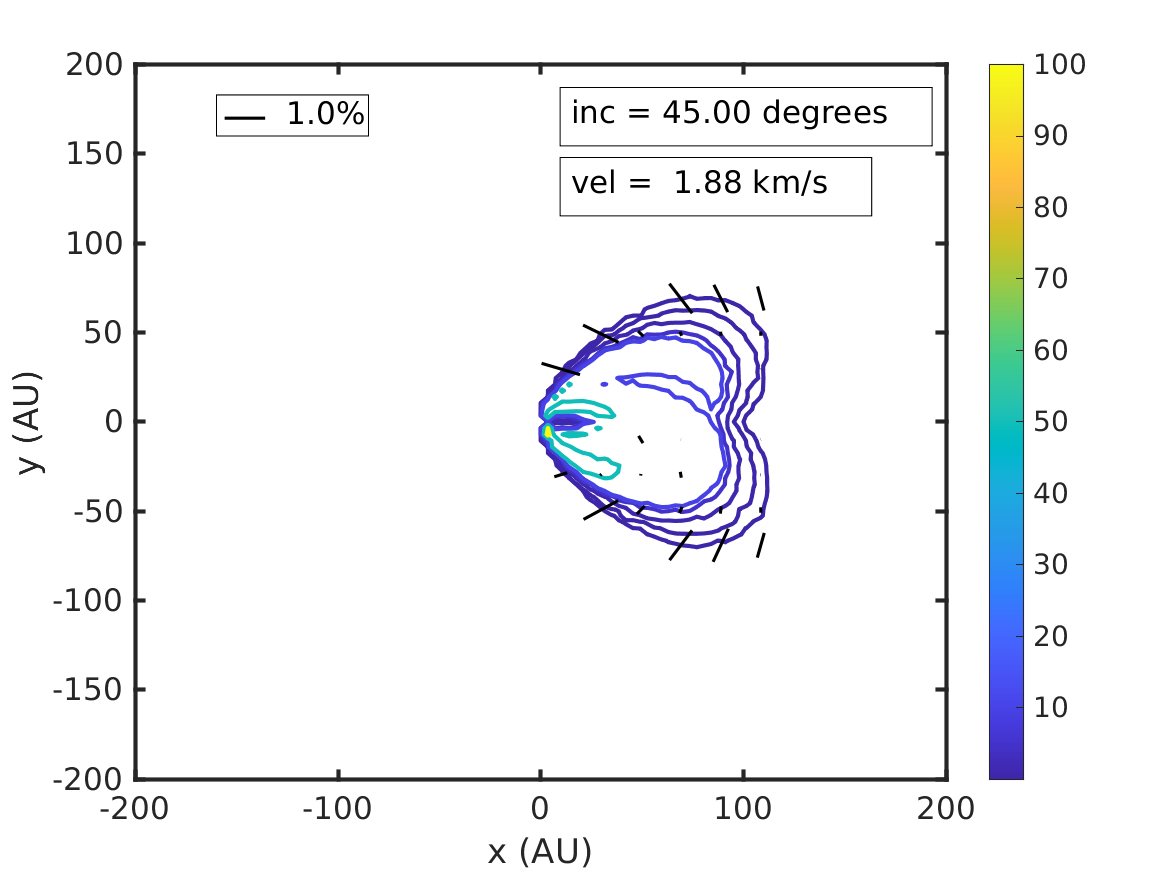}
       \caption{}
    \end{subfigure}
  \caption{Total intensity (in Kelvin) of a protoplanetary disk overlaid with polarization obtained from PORTAL simulations using a toroidal magnetic field. Polarization vector lengths are truncanted above $1 \%$. Subfigures (a)-(e) are of the $v=0$ km/s channel at inclination $0$, $30$ and $45$, and $60$ and $90$ degrees inclination. Subfigures (f)-(h) are of the velocity channels $v=0.62$, $v=1.25$, and $v=1.88$ km/s at an inclination of $45$ degrees.}
\end{figure*}

\begin{figure*}
    \centering
    \begin{subfigure}[b]{0.32\textwidth}
       \includegraphics[width=\textwidth]{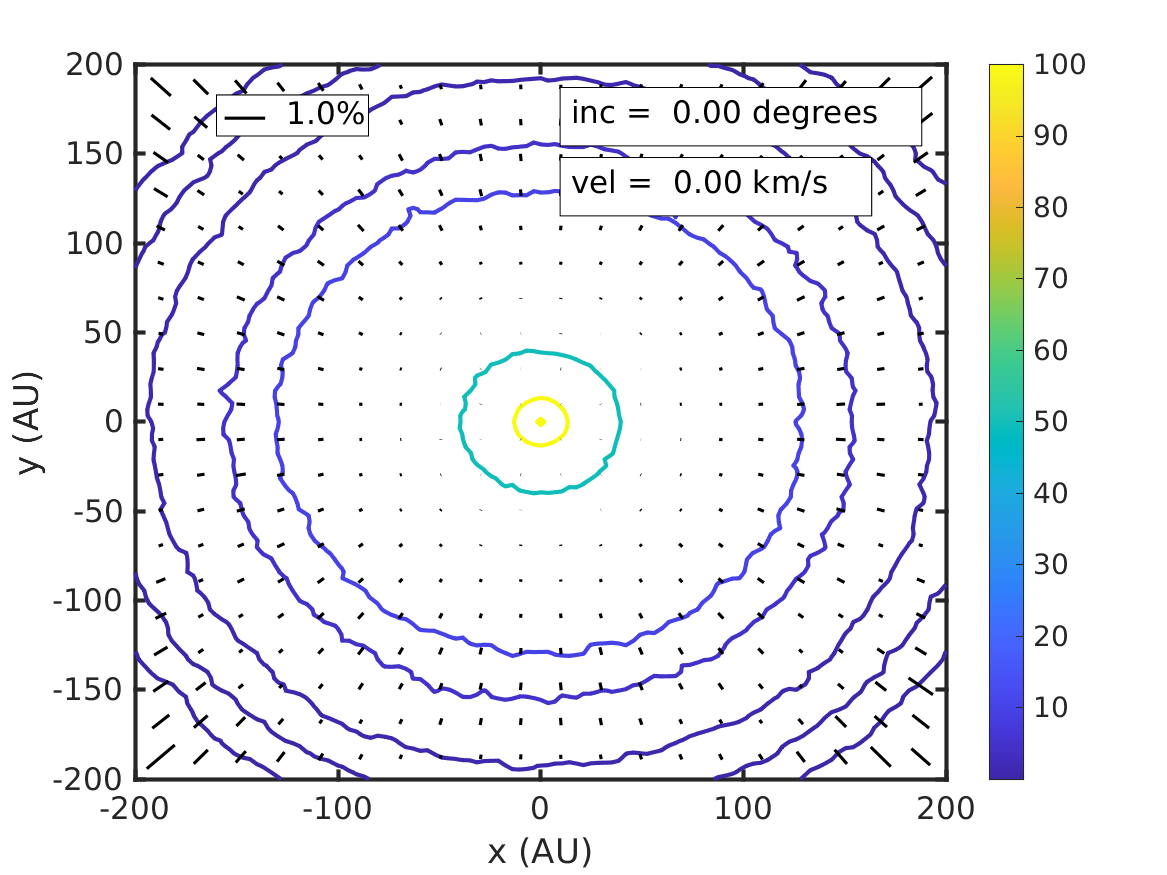}
       \caption{}
    \end{subfigure}
    ~
    \begin{subfigure}[b]{0.32\textwidth}
       \includegraphics[width=\textwidth]{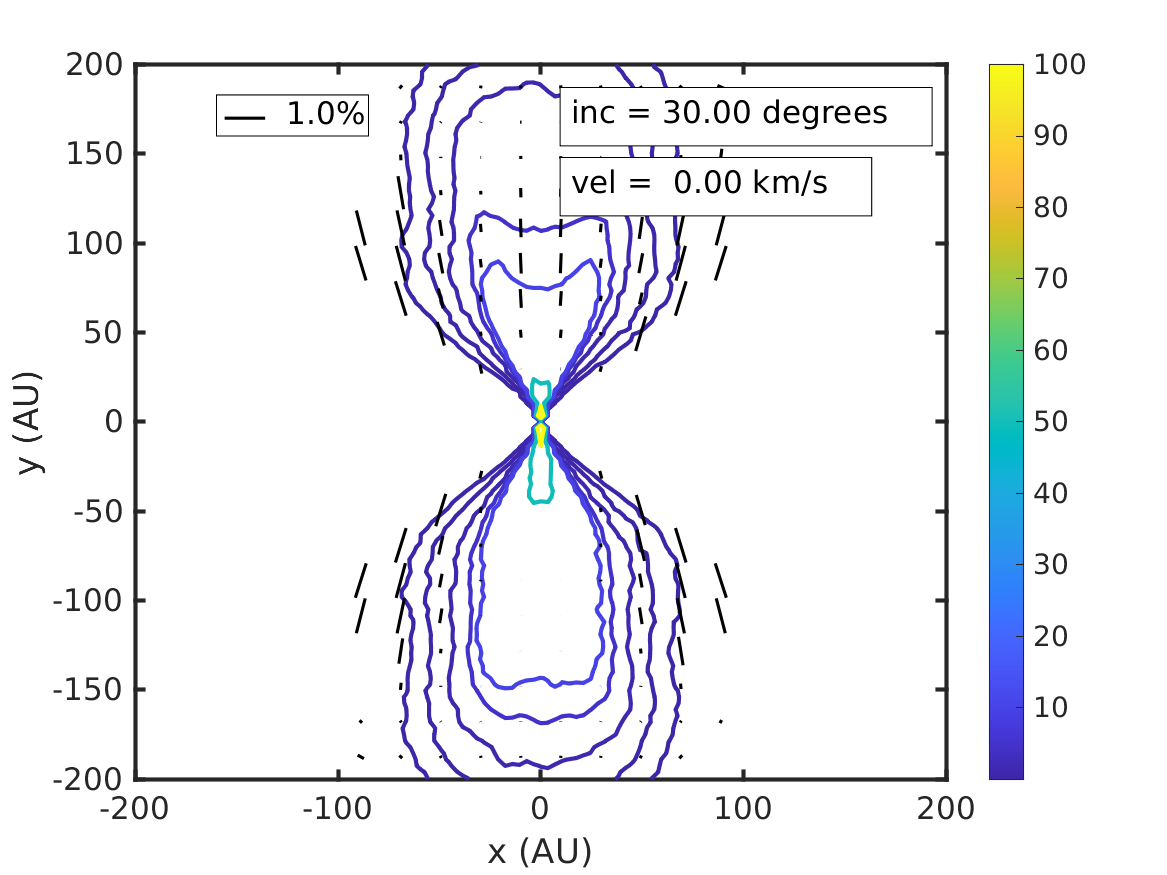}
       \caption{}
    \end{subfigure}
     ~
    \begin{subfigure}[b]{0.32\textwidth}
       \includegraphics[width=\textwidth]{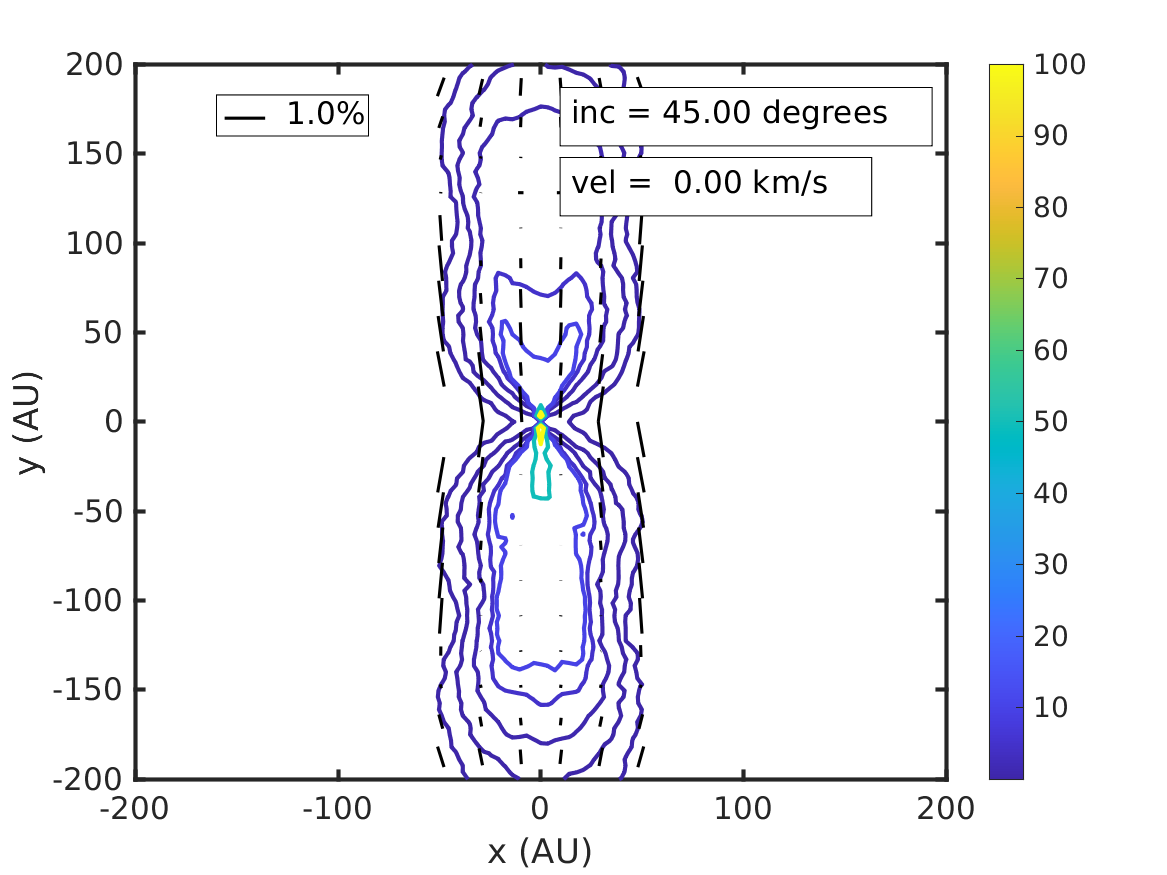}
      \caption{}
    \end{subfigure}
     ~
    \begin{subfigure}[b]{0.32\textwidth}
       \includegraphics[width=\textwidth]{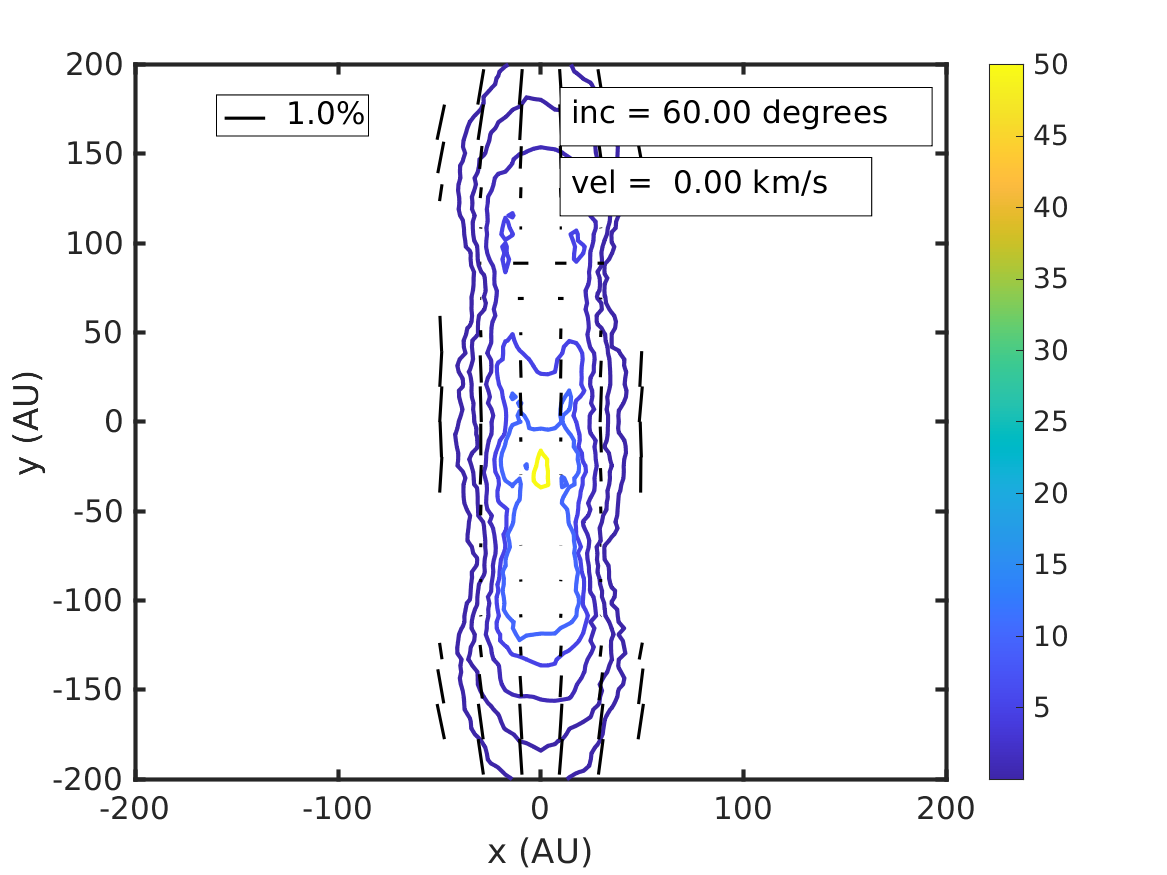}
       \caption{}
    \end{subfigure}
    ~
    \begin{subfigure}[b]{0.32\textwidth}
       \includegraphics[width=\textwidth]{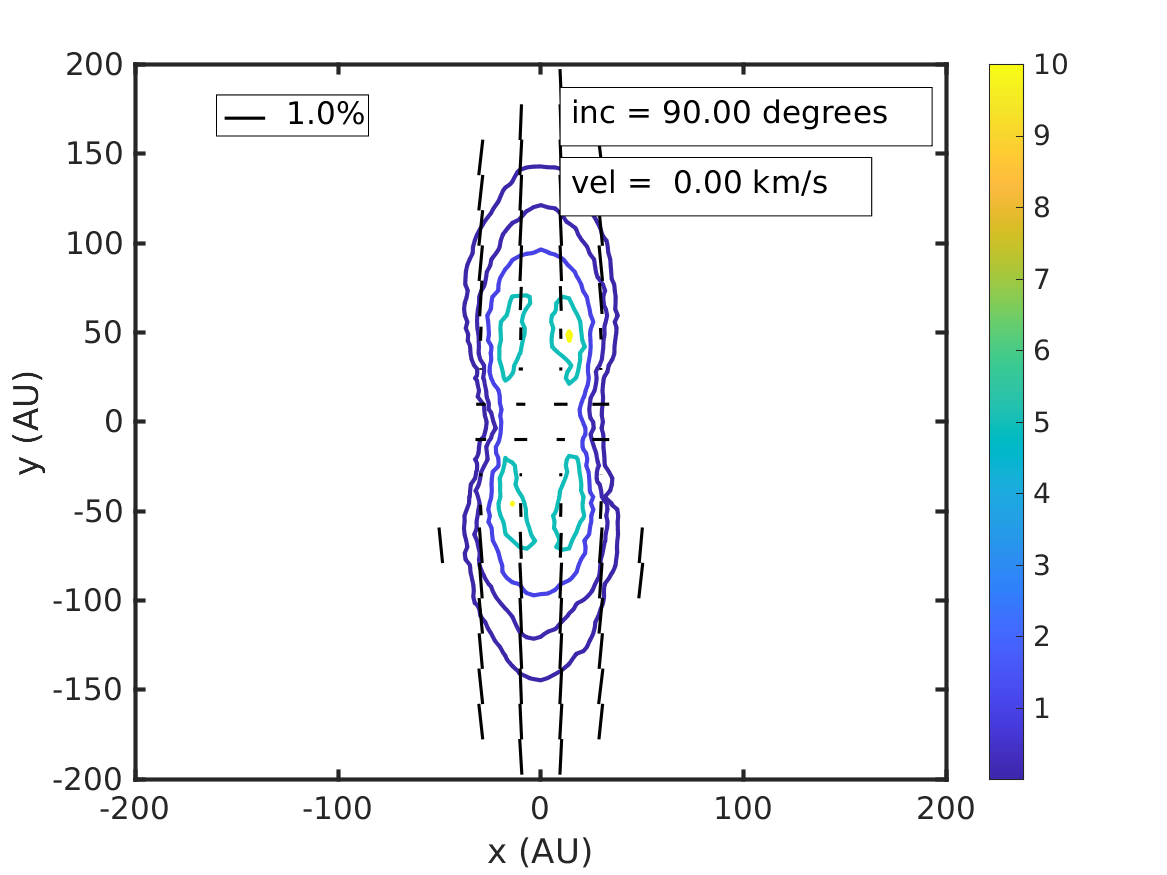}
       \caption{}
    \end{subfigure}

    \begin{subfigure}[b]{0.32\textwidth}
       \includegraphics[width=\textwidth]{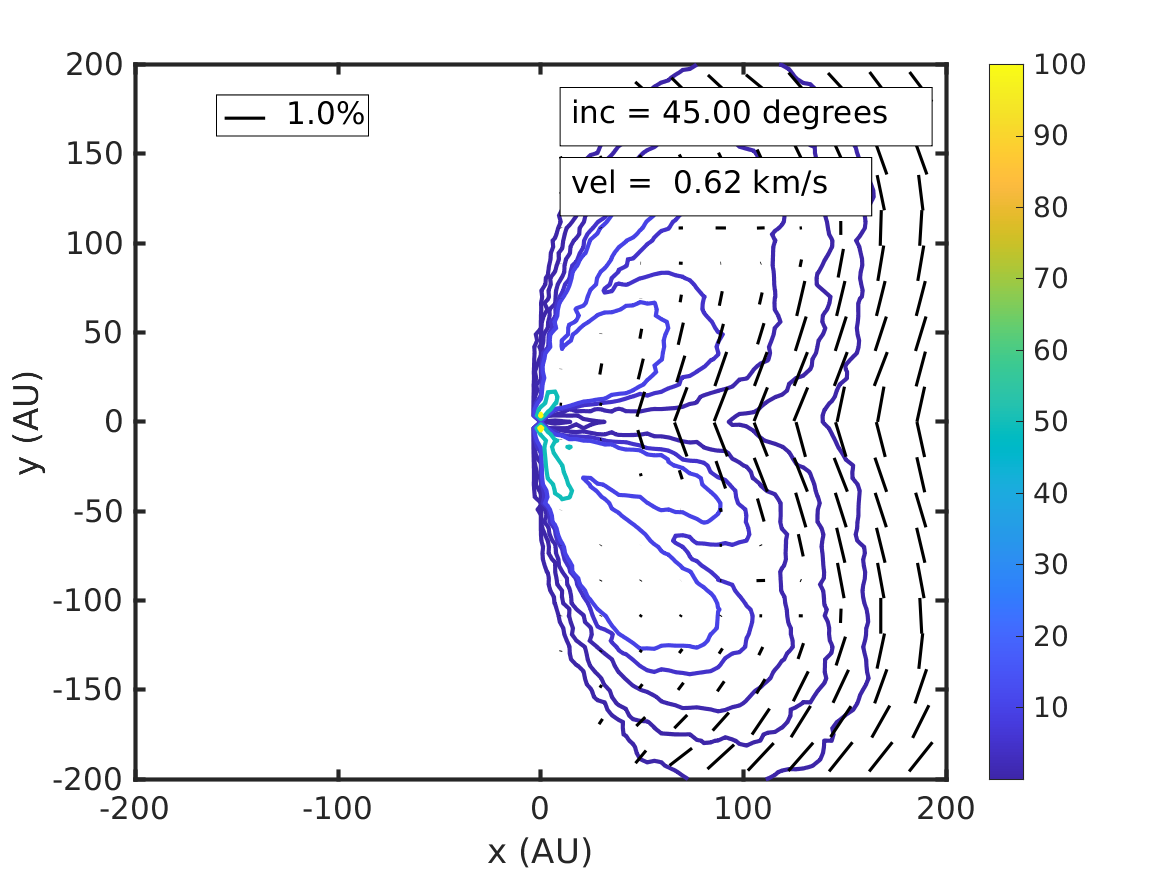}
      \caption{}
    \end{subfigure}
     ~
    \begin{subfigure}[b]{0.32\textwidth}
       \includegraphics[width=\textwidth]{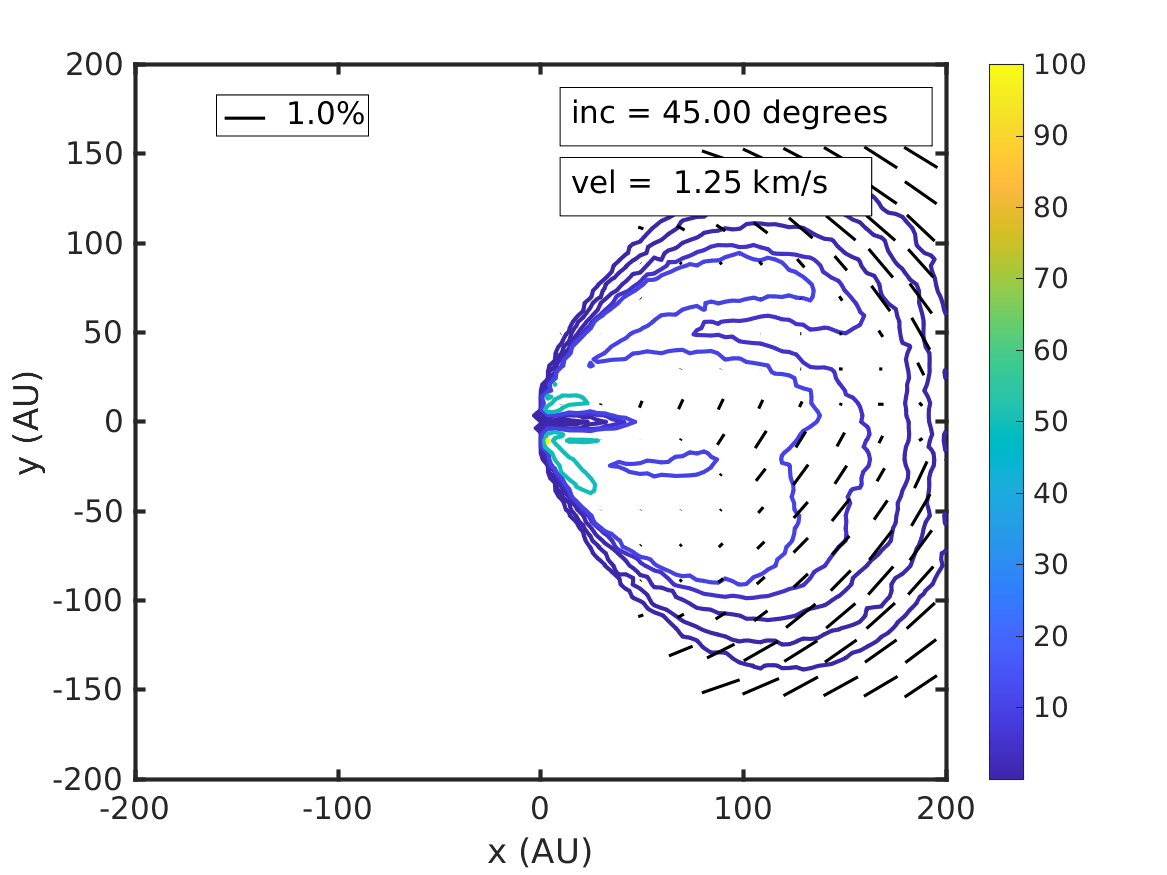}
       \caption{}
    \end{subfigure}
    ~
    \begin{subfigure}[b]{0.32\textwidth}
       \includegraphics[width=\textwidth]{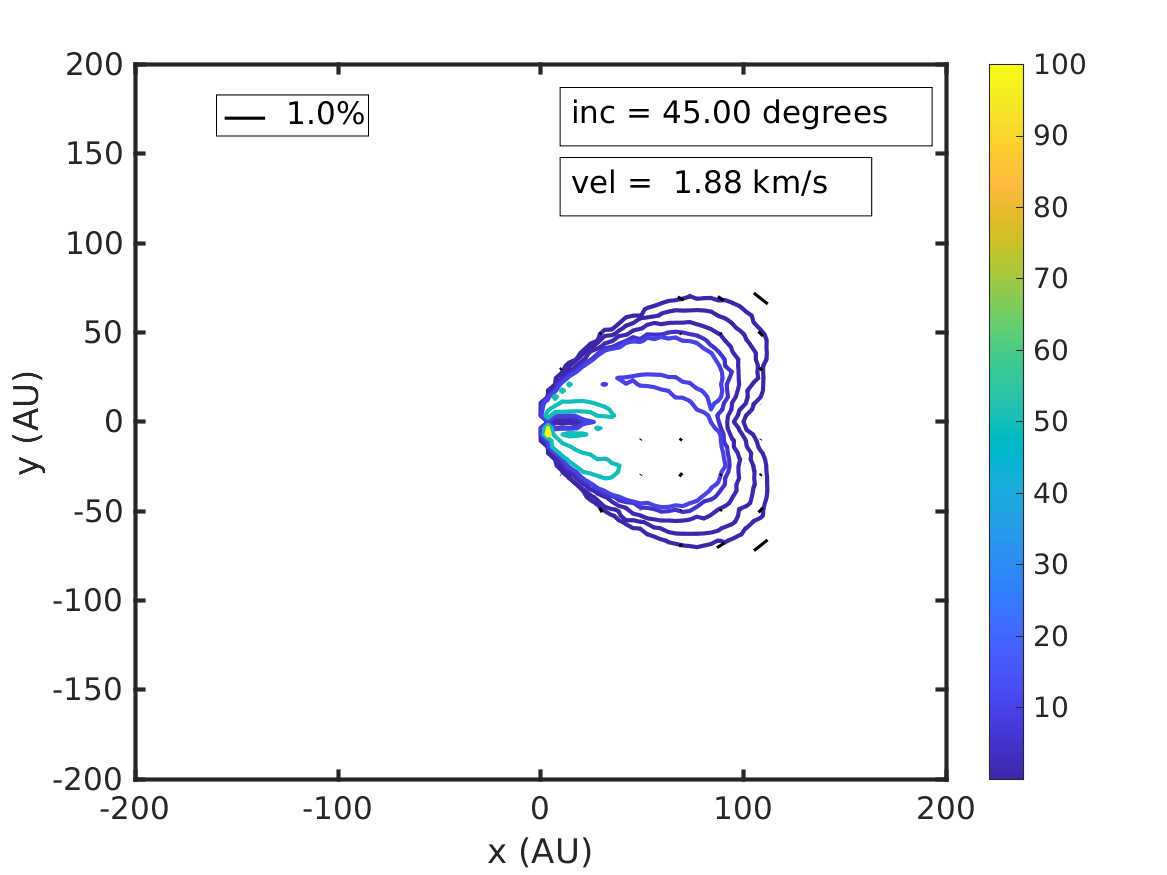}
       \caption{}
    \end{subfigure}
  \caption{Total intensity (in Kelvin) of a protoplanetary disk overlaid with the polarization obtained from PORTAL simulations using a poloidal magnetic field. Polarization vector lengths are truncanted above $1 \%$. Subfigures (a)-(e) are of the $v=0$ km/s channel at inclination $0$, $30$ and $45$, and $60$ and $90$ degrees inclination. Subfigures (f)-(h) are of the velocity channels $v=0.62$, $v=1.25$, and $v=1.88$ km/s at an inclination of $45$ degrees.}
\end{figure*}

\begin{figure*}
    \centering
    \begin{subfigure}[b]{0.32\textwidth}
       \includegraphics[width=\textwidth]{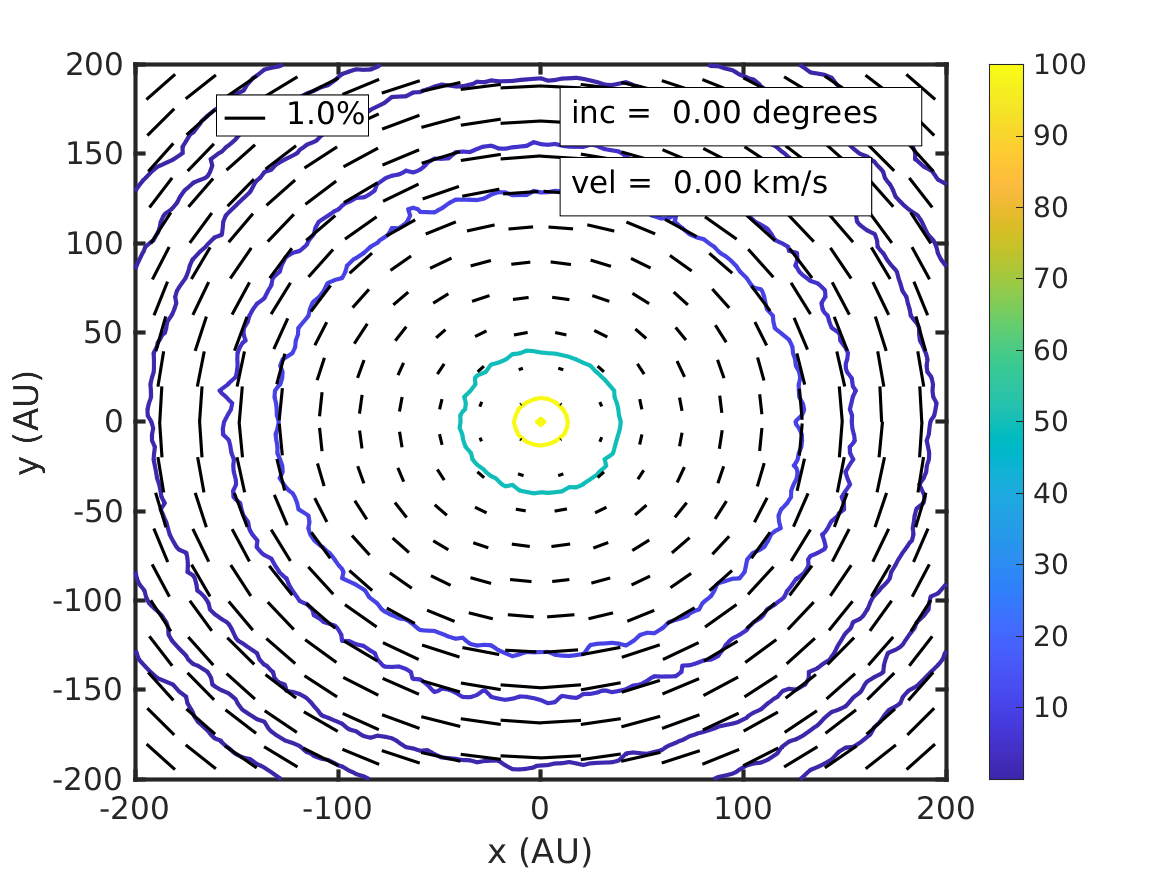}
       \caption{}
    \end{subfigure}
    ~
    \begin{subfigure}[b]{0.32\textwidth}
       \includegraphics[width=\textwidth]{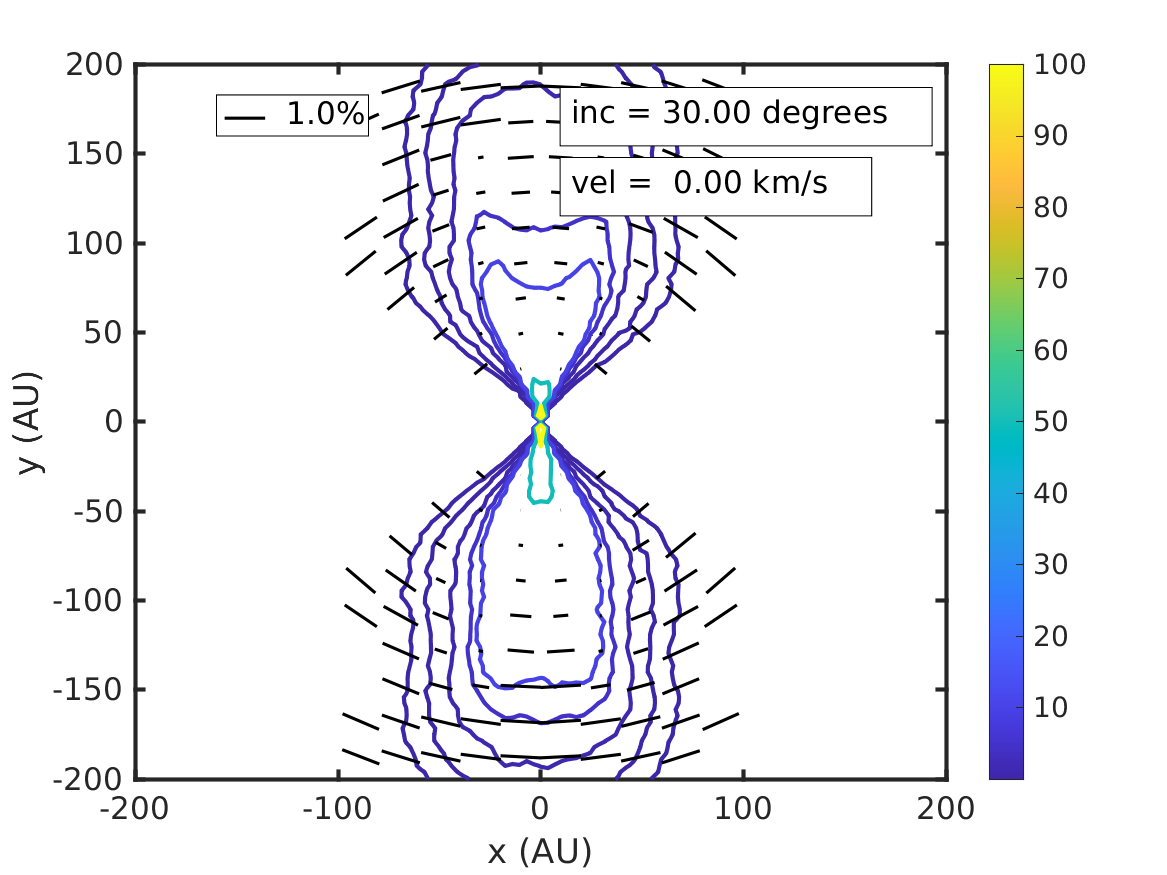}
       \caption{}
    \end{subfigure}
     ~
    \begin{subfigure}[b]{0.32\textwidth}   
       \includegraphics[width=\textwidth]{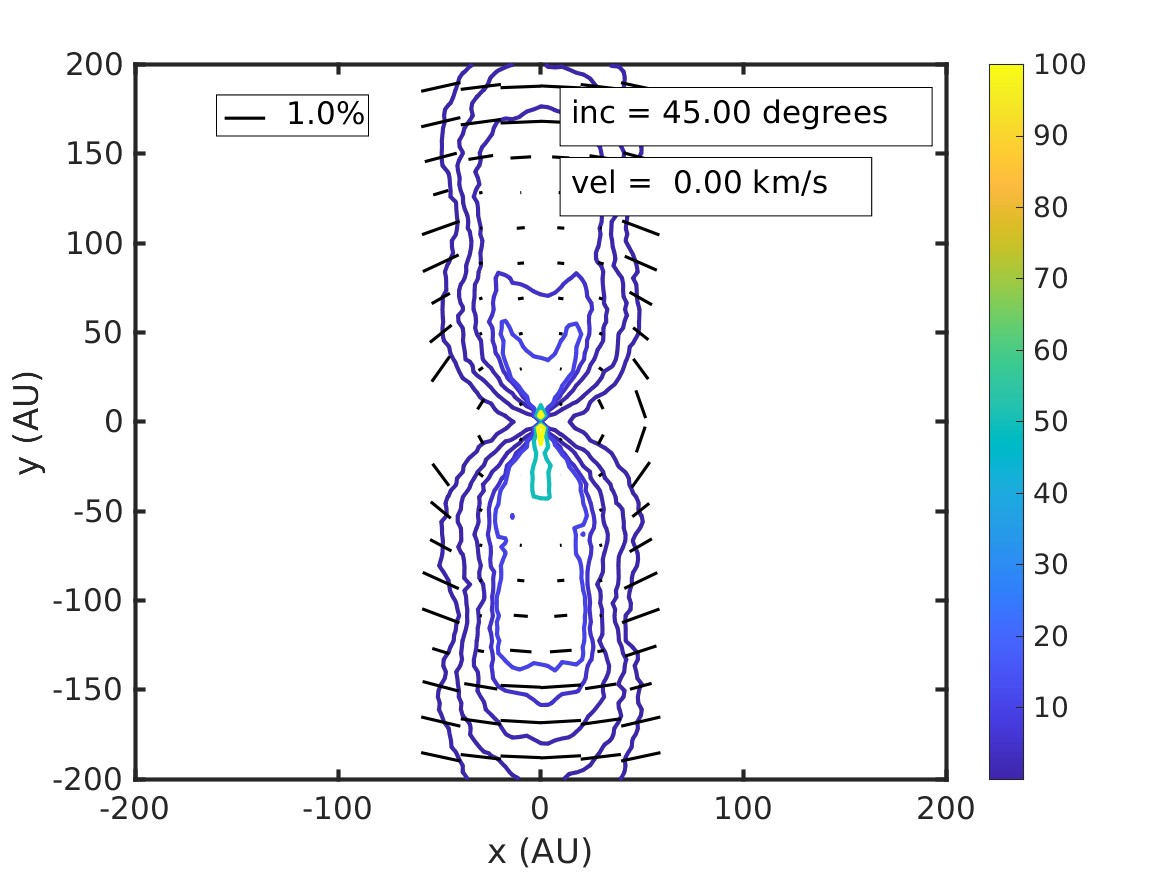}
      \caption{}
    \end{subfigure}
     ~
    \begin{subfigure}[b]{0.32\textwidth}
       \includegraphics[width=\textwidth]{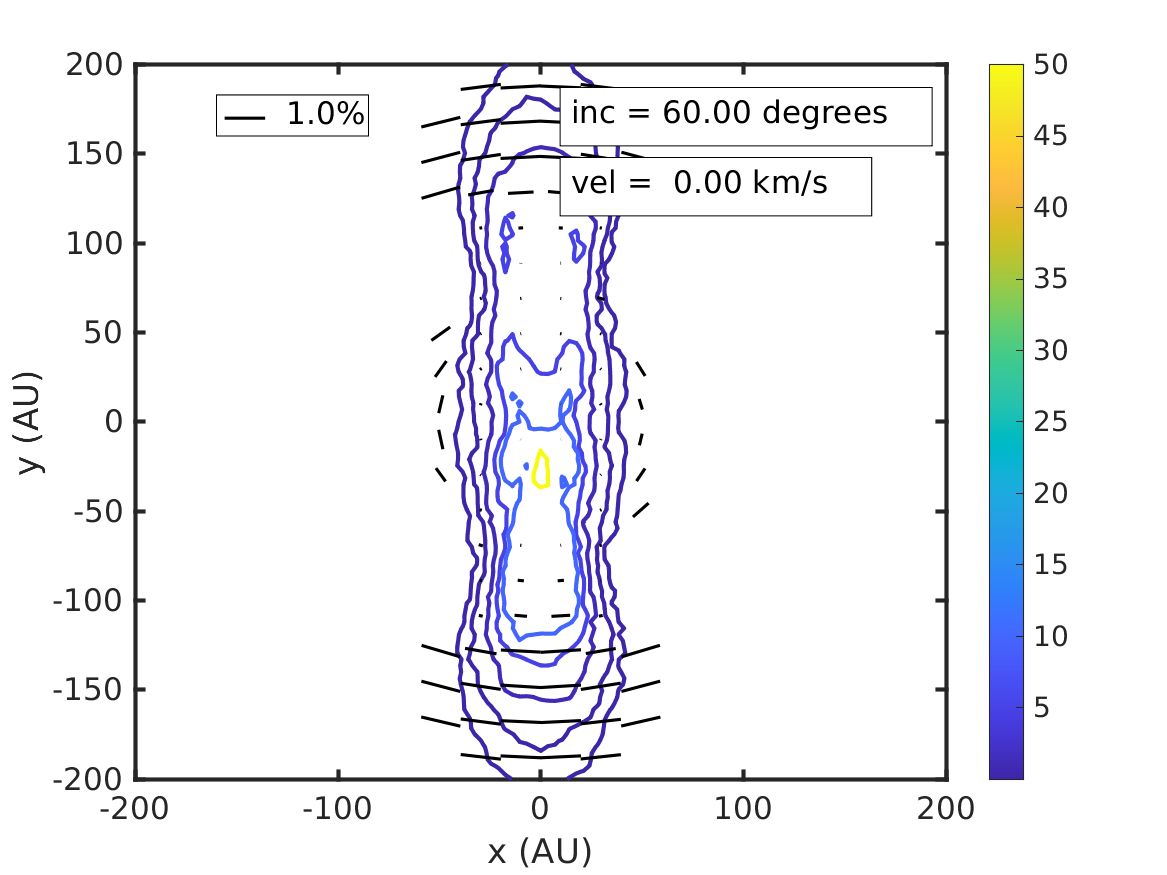}
       \caption{}
    \end{subfigure}
    ~
    \begin{subfigure}[b]{0.32\textwidth}
       \includegraphics[width=\textwidth]{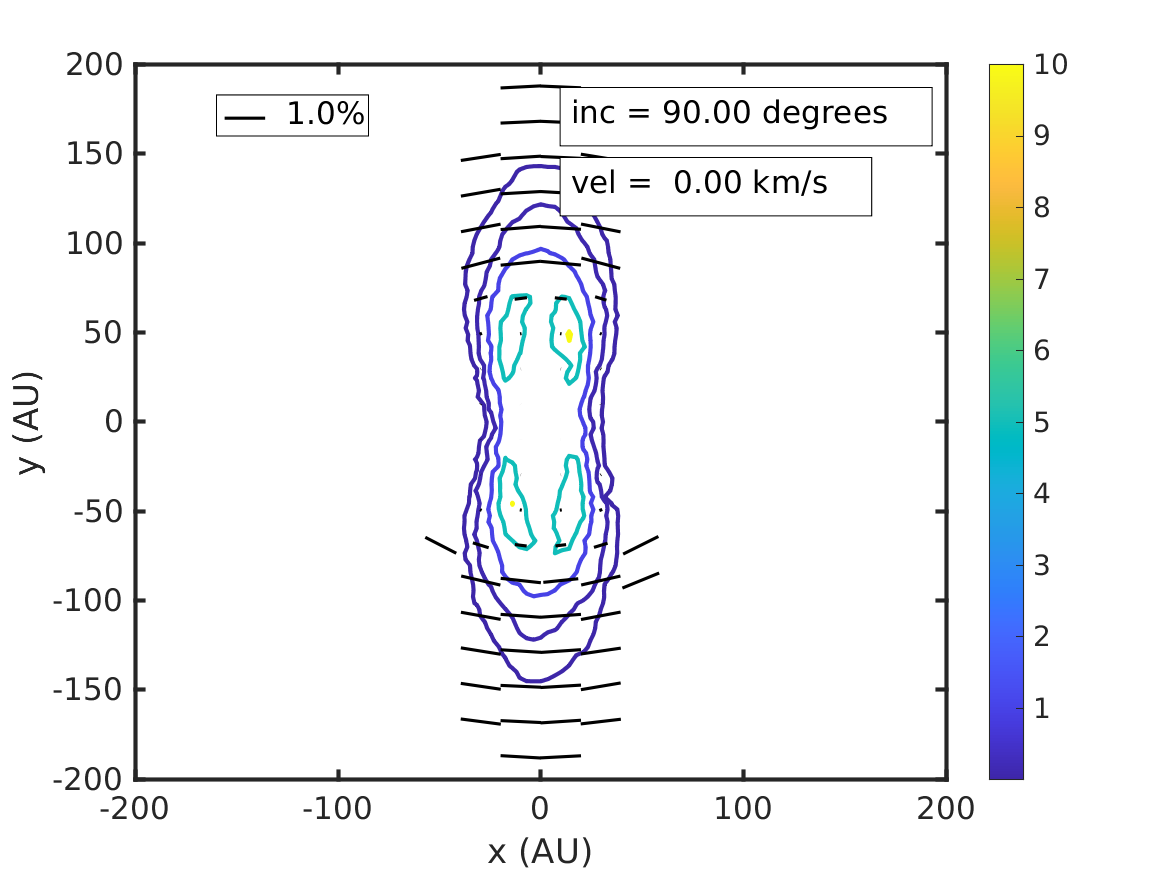}
       \caption{}
    \end{subfigure}

    \begin{subfigure}[b]{0.32\textwidth}
       \includegraphics[width=\textwidth]{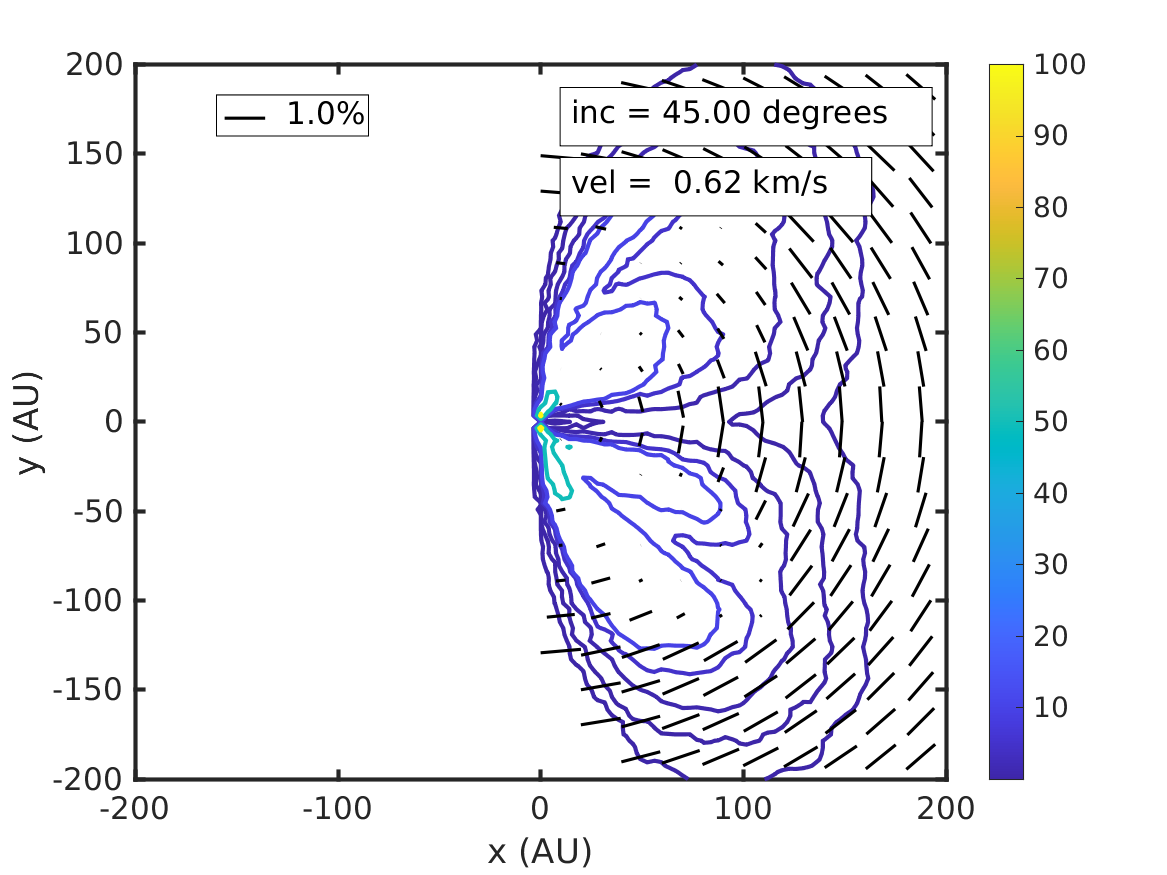}
      \caption{}    
    \end{subfigure}
     ~
    \begin{subfigure}[b]{0.32\textwidth}
       \includegraphics[width=\textwidth]{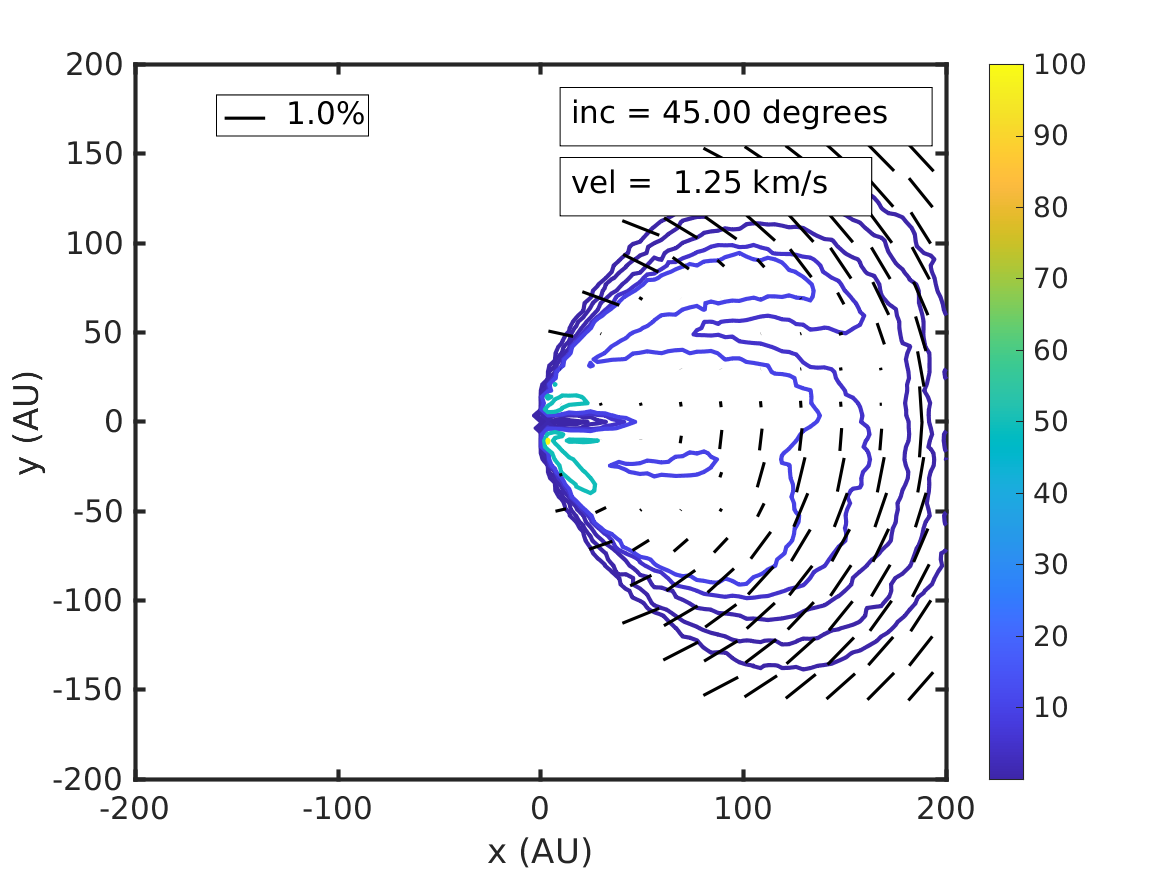}
       \caption{}
    \end{subfigure}
    ~
    \begin{subfigure}[b]{0.32\textwidth}
       \includegraphics[width=\textwidth]{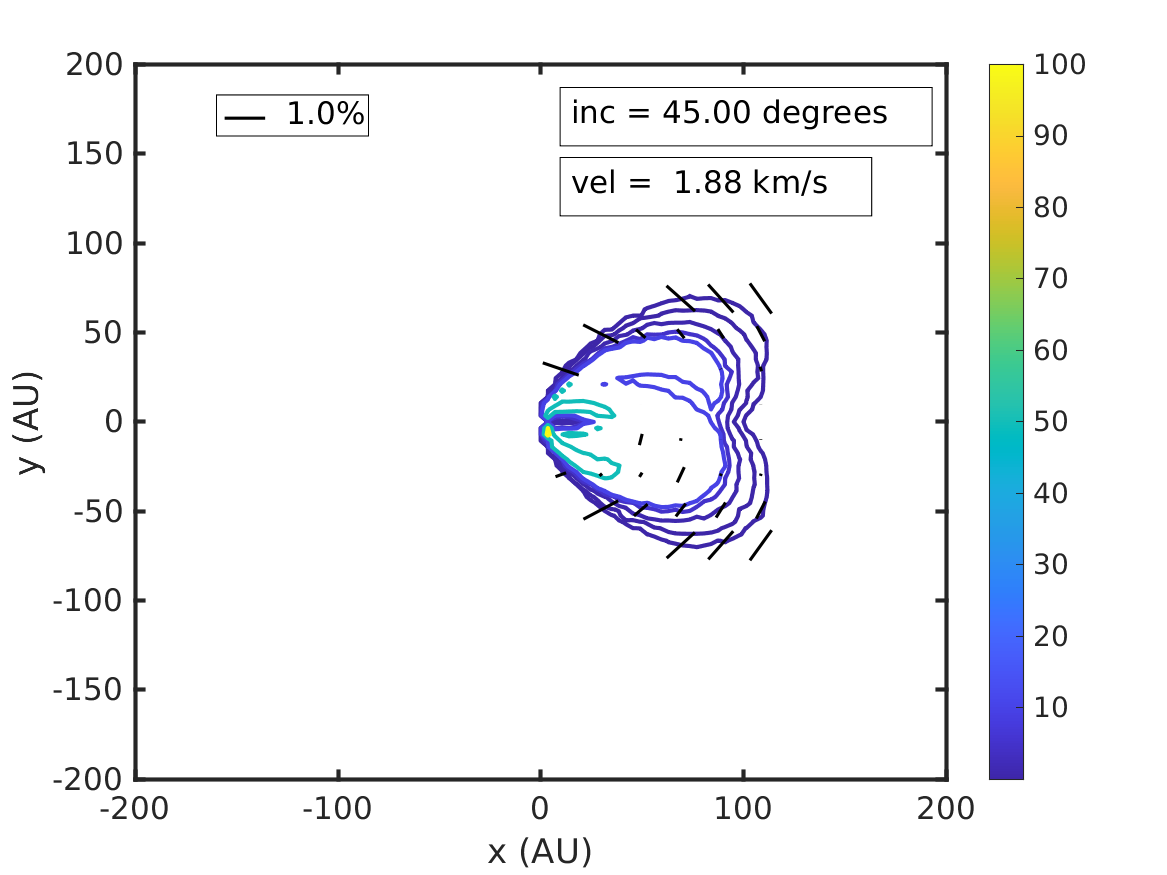}
       \caption{}
    \end{subfigure}
  \caption{Total intensity (in Kelvin) of a protoplanetary disk overlaid with the polarization obtained from PORTAL simulations using a radial magnetic field. Polarization vector lengths are truncanted above $1 \%$. Subfigures (a)-(e) are of the $v=0$ km/s channel at inclination $0$, $30$ and $45$, and $60$ and $90$ degrees inclination. Subfigures (f)-(h) are of the velocity channels $v=0.62$, $v=1.25$, and $v=1.88$ km/s at an inclination of $45$ degrees.}
\end{figure*}
\end{appendix}

\end{document}